\documentclass[12pt]{article}
\usepackage{a4,amsmath,epsfig,cite}
\newcommand{\Section}[1]{Section~\ref{#1}}
\newcommand{\Figure}[1]{Fig.~\!\ref{#1}}
\newcommand{\Expression}[1]{(\ref{#1})}
\newcommand{\Equation}[1]{Eq.~\!(\ref{#1})}
\newcommand{\eg}{{\it eg.}}

\newcommand{\JJ}{J}
\newcommand{\pp}{p}
\newcommand{\qq}{q}
\newcommand{\psl}{p\hspace{-5.5pt}/}
\newcommand{\Jsl}{\JJ\hspace{-7pt}/}
\newcommand{\ksl}{k\hspace{-6pt}/}
\newcommand{\esl}{e\hspace{-5.5pt}/}
\newcommand{\Ta}{T^{a}}
\newcommand{\kA}{k_1}
\newcommand{\eA}{e_1}
\newcommand{\ksA}{\ksl_1}
\newcommand{\esA}{\esl_1}
\newcommand{\Tb}{T^{b}}
\newcommand{\kB}{k_2}
\newcommand{\eB}{e_2}
\newcommand{\ksB}{\ksl_2}
\newcommand{\esB}{\esl_2}
\newcommand{\kAB}{k_{1+2}}
\newcommand{\ksAB}{\ksl_{1+2}}
\newcommand{\eAB}{e_{1+2}}
\newcommand{\esAB}{\esl_{1+2}}
\newcommand{\fAB}{\hat{e}_{1+2}}
\newcommand{\fsAB}{\hat{e}\hspace{-5.5pt}/_{1+2}}
\newcommand{\inp}[2]{#1\!\cdot\!#2}
\newcommand{\Icom}{I^{[1,2]}}
\newcommand{\Iant}{I^{\{1,2\}}}
\newcommand{\IAB}{I^{(1,2)}}
\newcommand{\IBA}{I^{(2,1)}}
\newcommand{\IA}{I^{(1)}}
\newcommand{\Ma}{\mathcal{M}^{a}}
\newcommand{\Mab}{\mathcal{M}^{a,b}}

\newcommand{\Imix}{I_{\mathrm{mix}}}
\newcommand{\Imixcom}{\Imix^{[1,2]}}
\newcommand{\Imixant}{\Imix^{\{1,2\}}}
\newcommand{\ImixAB}{\Imix^{(1,2)}}
\newcommand{\ImixBA}{\Imix^{(2,1)}}
\newcommand{\SU}{\mathrm{SU}}
\newcommand{\Tr}{\mathrm{Tr}}
\newcommand{\ncol}{N_{\mathrm{c}}}
\newcommand{\calM}{\mathcal{M}}
\newcommand{\imag}{\mathrm{i}}
\newcommand{\formula}[1]{\mbox{#1}}
\newcommand{\fA}{a_1}
\newcommand{\fB}{a_2}
\newcommand{\fsA}{a\hspace{-6pt}/_1}
\newcommand{\fsB}{a\hspace{-6pt}/_2}
\begin{document}
%
\begin{flushright}
{\normalsize CERN-PH-TH/2008-023}\\
{\normalsize IFJPAN-IV-2007-12} \\
{\normalsize 6 March 2008}
\end{flushright}
\vspace{1\baselineskip}
\begin{center}
%
{\bf\Large Gauge invariant sub-structures of tree-level}\\\vspace{0.2\baselineskip}
{\bf\Large double-emission exact QCD spin amplitudes%
}

\vspace{2\baselineskip}
{\bf A.~van Hameren\footnote{{\tt hameren@ifj.edu.pl}}}

{\it IFJ-PAN, ul.~Radzikowskiego 152, Krak\'ow, Poland}

\vspace{1\baselineskip}
{and}

\vspace{1\baselineskip}
{\bf Z.~W\c{a}s\footnote{{\tt Zbigniew.Was@cern.ch}\\
{\scriptsize
This work is partly supported by the EU grant mTkd-CT-2004-510126
in partnership with the CERN Physics Department,
by  RTN European Programme, MRTN-CT-2006-035505 (HEPTOOLS, Tools and Precision Calculations for Physics Discoveries at  Colliders)
 and by the Polish Ministry of
Scientific Research and Information Technology grant No 620/E-77/6.PRUE/DIE 188/
2005-2008.
}
}}

{\it IFJ-PAN, ul.~Radzikowskiego 152, Krak\'ow, Poland}\\
{\it and}\\
{\it CERN, 1211 Geneva 23, Switzerland}

\vspace{2\baselineskip}

\vspace{2\baselineskip}
{\bf Abstract}\\\vspace{0.5\baselineskip}
\parbox{0.8\linewidth}{\small\hspace{15pt}%
In this note we discuss possible separations of exact, massive, tree-level spin amplitudes into gauge invariant parts.
We concentrate our attention on processes involving two quarks entering a color-neutral current and, thanks to the QCD interactions, two extra external gluons.
We will search for forms compatible with parton shower languages, without applying approximations or restrictions on phase space regions.
Special emphasis will be put on the isolation of parts necessary for the construction of evolution kernels for individual splittings and to some degree for the running coupling constant as well.
Our aim is to better understand the environment necessary to optimally match hard matrix elements with partons shower algorithms.
To avoid complications and ambiguities related to regularization schemes, we ignore, at this point, virtual corrections.
%
%
%
Our representation is quite universal: any color-neutral current can be used, in particular our approach is not restricted to vector currents only.
}
\end{center}
\vspace{1\baselineskip}


\section{Introduction}
One of the essential steps in the construction of any algorithm for multi-particle final states is the appropriate analysis of the phase space parametrization.
In the PHOTOS Monte Carlo \cite{Barberio:1994qi} for multi-photon production, such an exact phase space parametrization is embodied in an iterative algorithm, the details of which are best described in \cite{Nanava:2006vv}, but the control
 of the relative size of sub-samples of distinct numbers of final state particles requires a precise knowledge of the matrix elements including virtual corrections.
In the KKMC Monte Carlo, the approach to phase space generation is different, but the necessity to control matrix elements and their truncation is again essential \cite{kkcpc:1999,Jadach:2000ir}.

Iterative procedures for parts of amplitudes, which are at the foundation of exponentiation \cite{Jadach:2000ir,Yennie:1961ad} and structure functions \cite{Altarelli:1977zs,Gribov:1972ri,Gorshkov:1966ht,Skrzypek:1992vk,RichterWas:1985gi} were exploited for the sake of use in KKMC Monte Carlo.
In particular the  description of the dominantly $s$-channel process $e^+e^- \to \nu_e \bar \nu_e \gamma \gamma $ where  $t$-channel $W$-exchange diagrams with gauge boson couplings contribute to
matrix elements provided an interesting example \cite{Was:2004ig}.
These studies were motivated by  practical reasons, but also pointed at quite astonishing properties of tree-level spin amplitudes, namely that they can be separated into gauge invariant parts in a semi-automated way, easy to apply in the Kleiss-Stirling methods \cite{Kleiss:1985yh,Jadach:1998wp}.

One could ask the question whether similar techniques can be used in QCD, whether they are  of any practical use, and in fact to which degree they were already included in previous work, such as for example at the foundation of parton shower algorithms (new or well established)
\cite{Bukhvostov:1985rn,Azimov:1986sf,Gustafson:1987rq,Andersson:1989ki,Marchesini:1987cf,Ellis:1986bv,Marchesini:1983bm,Nagy:2008ns,Nagy:2007ty} or for other, fundamental or  phenomenological applications \cite{Bern:2004ba,Rodrigo:2005eu,Badger:2005jv,Andersen:2007mp,JezuitaDabrowska:2002jw,Somogyi:2006da,Berends:1987me,Berends:1988yn}.
The common idea between all of these papers is to separate approximate or exact results into parts, often with the help of iteration.
We will not elaborate on these points here, this broad subject by far exceeds the purpose of the present paper.
We only want to mention the existence of possible limitations in such strategies, if applied to parton shower applications \cite{Kleiss:1990jv}.
Our study will be purely technical, and we will also ignore any possible complications related to confinement for example.
As was done for QED, we will study possible ways of separating amplitudes, but now for the process of two quarks and two gluons entering a color-neutral current.
First, we try to identify sub-structures of the amplitude which would be useful for possible applications.
Then we would like to verify to which degree they can be used to localize parts of the amplitude related to \eg\ evolution kernels and the running coupling constant.
We expect such expressions to be identifiable at least in approximations valid in certain regions of phase space, dominant for specific purposes, like in applications using \formula{$p_T$} ordered phase space, but we hope to localize them
partly already at the level of exact amplitudes.

However, we also expect that the choice of what we call {\em subtraction terms\/} will be more ambiguous than in the QED case.
It concerns the terms which are subtracted at one place of an expression, and added at another one.
For QED, these terms could be constructed from parts of lower-order amplitudes, and resulted in a semi-automated method of organizing amplitudes.
There, the most singular parts of \eg\ double photon bremsstrahlung amplitudes could be deduced from amplitudes for single photon emission.
In fact, those properties are known since long and necessary for exponentiation.
In the calculation of the so-called \formula{$\beta_2$} for double-photon emission, the most singular parts are {\em subtracted\/} from the amplitude.
This is not only convenient but also necessary, since those terms are already included, in a rigorous way, through the exponentiation of lower order amplitudes.
In our present study, it will not be possible to construct the subtraction terms in such an unambiguous way.
Our expressions concern the amplitude for the process \formula{$q\bar{q}\to\JJ{}gg$}, but because they are exact, the incoming and outgoing particles can be easily interchanged to describe for example the process \formula{$qg\to\JJ{}gq$}.
The amplitudes for the two processes are basically the same, but the lower order processes, for which they may be considered as real-emission corrections, are different.
This clearly points at ambiguities if the form of lower order amplitudes is  to be used for the definition of subtraction terms.
Obviously, by inspection of the equations for the evolution of the parton showers or PDFs, both of the two processes are potentially of similar phenomenological importance.
%


Our paper is organized as follows.
In \Section{SecNotation} we present our notation and general strategy to organize the amplitude.
The treatment of different possibilities for this organization concerning the color part of the amplitude is distributed over \Section{SecCommAnti} and \Section{SecSingProd}.
We start with a formulation close to the one used in previous papers on QED, moving consecutively towards shorter and more natural forms for QCD applications.
In \Section{SecCases} we explore certain limits, in which some parts of the amplitude can be dropped out.
The summary in \Section{SecSummary} closes the paper.
%

\section{Notation and general strategy\label{SecNotation}}
%
The exact spin amplitude for the process \formula{$q\bar{q}\to\JJ gg$} can be written as an expansion in combinations of the \formula{$\SU(\ncol)$}-generators, for example the combinations \formula{$\{\Ta,\Tb\}$} and \formula{$[\Ta,\Tb]$}.
Here, \formula{$\Ta$} is the color generator associated with gluon number $1$, which has momentum \formula{$\kA$} and polarization vector \formula{$\eA$}, and \formula{$\Tb$} is the color generator associated with gluon number $2$, which has momentum \formula{$\kB$} and polarization vector \formula{$\eB$}.
Another option would be to use, for example, the combinations \formula{$\Ta\Tb$} and \formula{$\Tb\Ta$}.
But let us stick to the former example for now, and return to this issue later.
The amplitude can then be expressed as
%
\begin{equation}
\Mab
=
\frac{1}{2}
\,\bar{v}(\pp)
\Big(
[\Ta,\Tb] \Icom + \{\Ta,\Tb\} \Iant
\Big)
u(\qq)
~,
\label{Eq271}
\end{equation}
%
where \formula{$\bar{v}(\pp)$} and \formula{$u(\qq)$} are the spinors associated with the anti-quark, which has momentum \formula{$\pp$}, and the quark, which has momentum \formula{$\qq$}, respectively. 
We do not specify the spin or color states for the quark fields, any choice can be used. This type of separation of the spin amplitude  into
gauge invariant parts is known \cite{Mangano:1988kk} and exploited since long time.
The main task is now to calculate the coefficients \formula{$\Icom$} and \formula{$\Iant$}.
They can be written as sums of terms, each of them containing factors consisting of scalar products of the aforementioned four-vectors and contractions of these four-vectors with the Dirac gamma-matrices.
Furthermore, each term will carry a factor \formula{$\Jsl$}, associated with the color-neutral current \formula{$\JJ$}.
We want to stress that all expressions we present are valid under the condition that the current \formula{$\Jsl$} is constructed from color neutral objects like \formula{$(v+a\gamma^5)$}, \formula{$\gamma^\mu$}, \formula{$\psl$}, \formula{$\ksA$}, etc.,
although our notation might seem to indicate stronger limitations.

\begin{figure}
\begin{center}
                       \epsfig{file=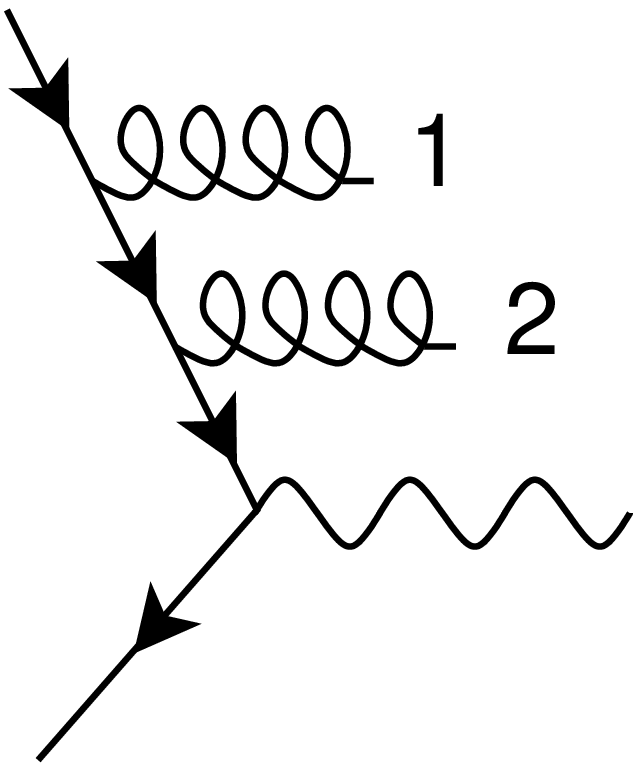,width=0.12\linewidth}
\hspace{0.05\linewidth}\epsfig{file=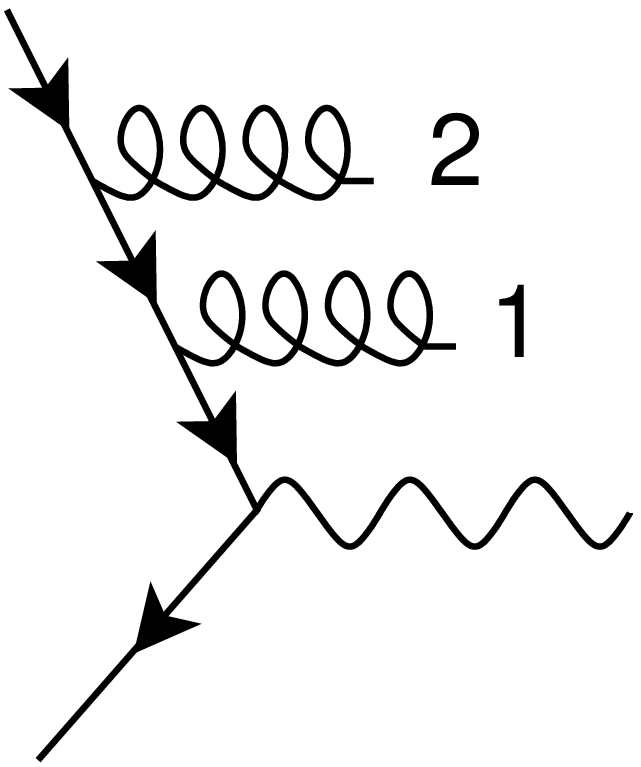,width=0.12\linewidth}
\hspace{0.05\linewidth}\epsfig{file=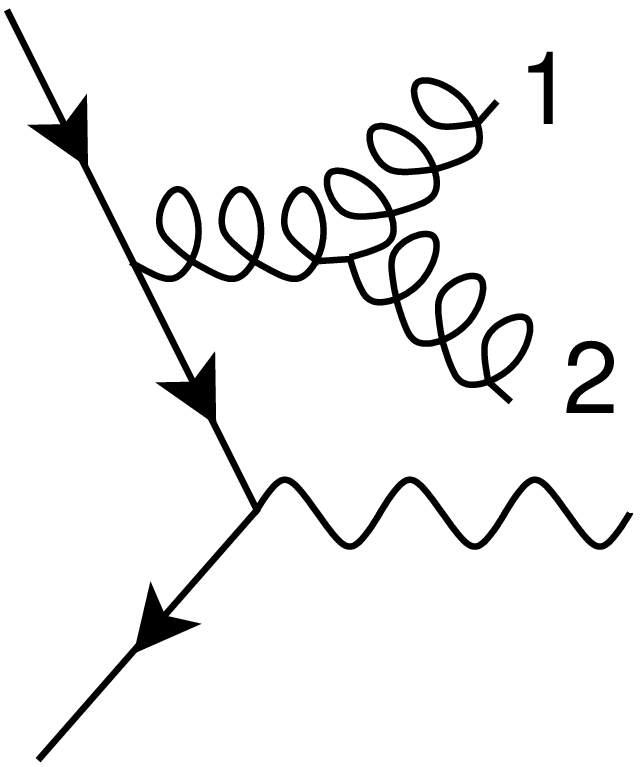,width=0.12\linewidth}
\hspace{0.05\linewidth}\epsfig{file=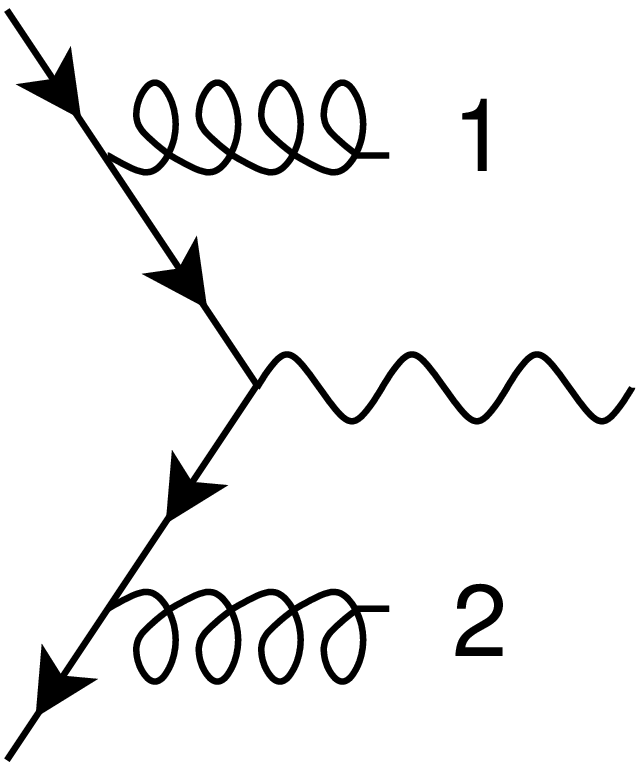,width=0.12\linewidth}
\\[3ex]
                       \epsfig{file=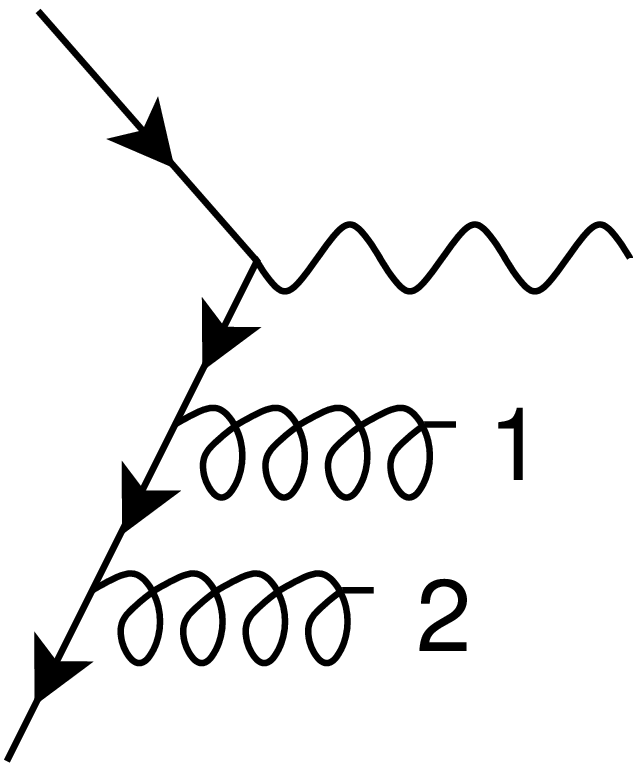,width=0.12\linewidth}
\hspace{0.05\linewidth}\epsfig{file=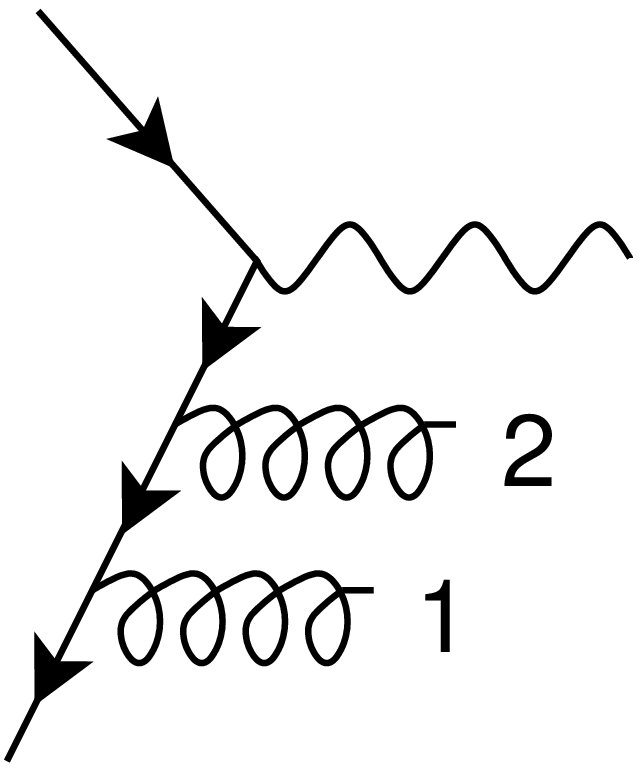,width=0.12\linewidth}
\hspace{0.05\linewidth}\epsfig{file=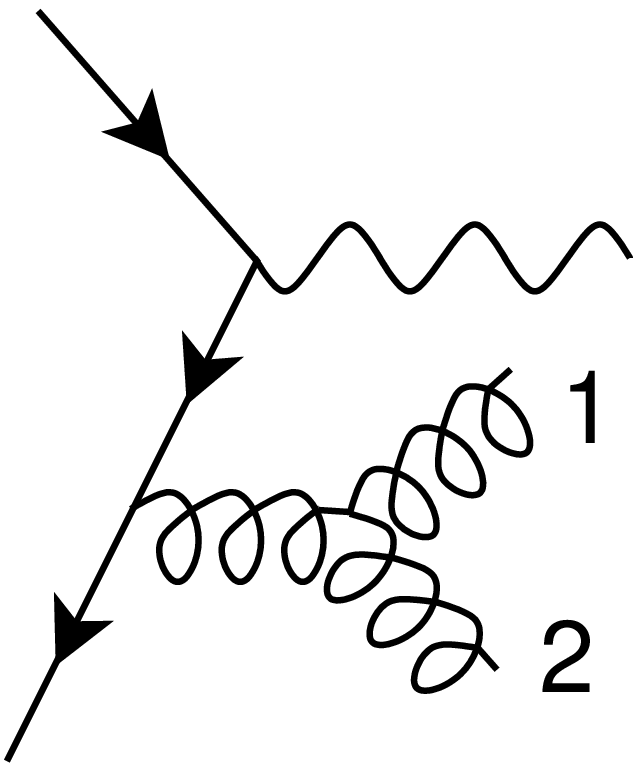,width=0.12\linewidth}
\hspace{0.05\linewidth}\epsfig{file=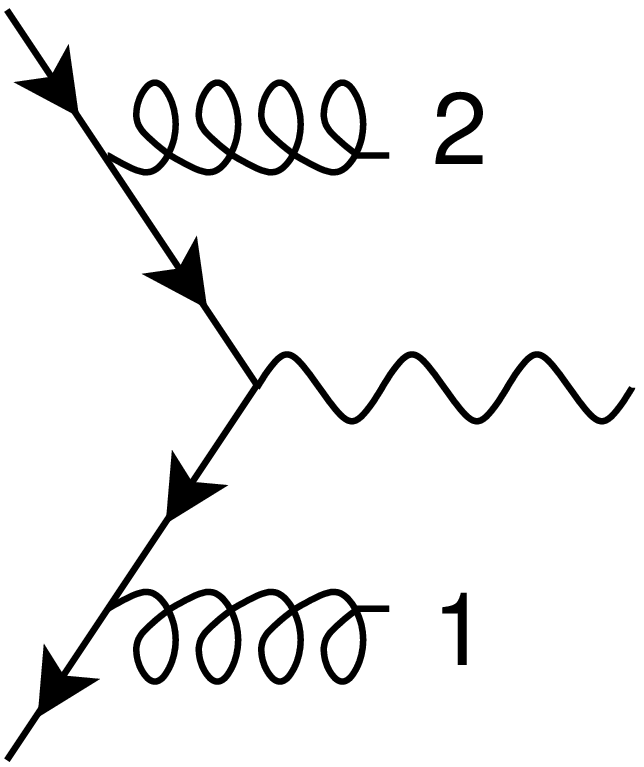,width=0.12\linewidth}
\caption{Feynman graphs for the process \formula{$q\bar{q}\to\JJ{}gg$}.\label{Fig1}}
\end{center}
\end{figure}

The first step in this calculation is the construction of the relevant graphs, depicted in \Figure{Fig1}, from the Feynman rules.
Next, we choose to replace explicit mass terms by fermion momenta using the Dirac equation.
It is not difficult to extract expressions of the type mentioned before for \formula{$\Icom$} and \formula{$\Iant$} at this point.
These expression will contain terms with products of zero, two or four gamma-matrix contractions (or ``slashed'' four-vectors), excluding the object \formula{$\Jsl$}.
Thanks to properties of the gamma-matrices, we can now choose how these products are ordered, at the expense of the appearance or disappearance of certain scalar products of four-momenta.
We choose to order them such that factors \formula{$\Jsl\esl_{1,2}$}, \formula{$\esl_{1,2}\Jsl$} or \formula{$\ksl_{1,2}\ksl_{2,1}$}, \formula{$\esl_{1,2}\esl_{2,1}$} are not present.
This is  for the convenience of possible use of the Kleiss-Stirling method for coding final expressions similarly as in \cite{kkcpc:1999}.

The game is now to compactify expressions, or organize them in an easily interpretable way, by factorizing certain sums of terms.
The most important guideline we use is that each term, consisting of factors which consist themselves of sums of terms, should be manifestly gauge invariant.
In this process, we allow for the introduction of what we call {\em subtraction terms\/}, that is terms subtracted at one place of the complete expression, and added at another place.
They can help to make certain factorizations possible, which look desirable but for which some terms appear to be missing.

Such subtraction terms were also used in our earlier QED-studies of the same type \cite{Was:2004ig}.
In fact, the terms were constructed from  parts of lower-order amplitudes and were an important element in the attempt to  define a semi-automated method to obtain expressions for spin amplitudes also used in Monte Carlo programs.
For the two-photon case, the following subtraction terms were used:
%
\begin{equation}
S^{\{1,2\}}_{1,\qq}
=
\frac{1}{2}
\,\Jsl
\bigg(
  \frac{\inp{\qq}{\eA}}{\inp{\qq}{\kA}} \;  \frac{\inp{\qq}{\eB}}{\inp{\qq}{\kA}+\inp{\qq}{\kB}}
+ \frac{\inp{\qq}{\eA}}{\inp{\qq}{\kA}+\inp{\qq}{\kB}}  \; \frac{\inp{\qq}{\eB}}{\inp{\qq}{\kB}}
\bigg)
~,
\label{a}
\end{equation}
\begin{equation}
S^{\{1,2\}}_{2,\qq}
=
\frac{1}{2}
\,\Jsl
\bigg(
  \frac{\ksB\esB}{\inp{\qq}{\kB}} \; \frac{\inp{\qq}{\eA}}{\inp{\qq}{\kA}}
+ \frac{\ksA\esA}{\inp{\qq}{\kA}} \; \frac{\inp{\qq}{\eB}}{\inp{\qq}{\kB}}
\bigg)
~,
\label{b}
\end{equation}
%
and similar terms with \formula{$\qq$} replaced by \formula{$\pp$}.
As mentioned before, these terms are constructed from 
parts of the one-photon amplitude, which is given by
%
\begin{equation}
\Ma
=
\bar{v}(\pp) \Ta\IA u(\qq)
~,
\label{firstOa}
\end{equation}
%
with
%
\begin{equation}
\IA
=
\Jsl
\bigg[
  \bigg( \frac{\inp{p}{e_1}}{\inp{p}{k_1}} - \frac{\inp{q}{e_1}}{\inp{q}{k_1}} \bigg)
\bigg]
- \frac{1}{2}\bigg[\frac{\esl_1\ksl_1}{\inp{p}{k_1}} \bigg] \Jsl
+ \frac{1}{2}\,\Jsl\bigg[\frac{\ksl_1\esl_1}{\inp{q}{k_1}}\bigg]
~.
\label{firstO}
\end{equation}
%
Note that each segment encapsulated by square brackets is manifestly gauge invariant.

As in the case of QED, we will use subtraction terms \Expression{a} and \Expression{b}.
In addition however, we will also use
%
\begin{equation}
S^{\{1,2\}}_{3,\qq}
=
-\frac{1}{2}
\,\Jsl
\,\frac{\inp{\kA}{\kB}}{\inp{\qq}{\kA}+\inp{\qq}{\kB}-\inp{\kA}{\kB}}
\; \frac{\inp{\kA}{\eB}\,\inp{\kB}{\eA}}{\inp{\kA}{\kB}\; \inp{\kA}{\kB}}
\label{c}
\end{equation}
%
and a similar term with \formula{$\qq$} replaced by \formula{$\pp$} as subtraction terms.
Note that, in contrast to \Expression{a} and \Expression{b}, subtraction term \Expression{c} is {\em not\/} constructed from elements present in the single bremsstrahlung amplitude given in \Equation{firstO}.
In particular, it introduces an artificial singularity \formula{$1/(\inp{\kA}{\kB})$} which would be worrisome for QED applications.
We, however, are not concerned with that, since it points at the singularity of an intermediate gluon virtuality, with which we have to deal anyway in a QCD amplitude.

We already mentioned the examples of an expansion in \formula{$[\Ta,\Tb]$} and \formula{$\{\Ta,\Tb\}$}, and an expansion in \formula{$\Ta\Tb$} and \formula{$\Tb\Ta$}.
We will deal with both cases in the next sections.
We start with the first alternative because the \formula{$\{\Ta,\Tb\}$}-part has exactly the same form as a two-photon amplitude in QED.
The second part, proportional to \formula{$[\Ta,\Tb]$}, is new in that respect.
Na\"\i{}vely, one would expect it can, to a large degree, be identified as a correction to the single gluon emission -- that is the case when this emission becomes virtual and decays into a pair of gluons.
For that part of the amplitude we will use subtraction terms quite similar to the ones discussed here.

The second alternative, of an expansion for the color contents in terms of \formula{$\Ta\Tb$} and \formula{$\Tb\Ta$}, is inspired by approximations exploited for example for parton showers and in particular dipole models.
Then, the gluons can be considered to appear consecutively (in one sense or another).

The most compact expressions for the amplitude will be obtained with a mixed choice, in which all four combinations \formula{$[\Ta,\Tb]$}, \formula{$\{\Ta,\Tb\}$}, \formula{$\Ta\Tb$} and \formula{$\Tb\Ta$} of color generators are used.
The coefficients for each term is not unique in this case of course, but we will be able to find a choice which seems to be optimal.

\section{QED-inspired picture\label{SecCommAnti}}
%
In this section, we investigate expressions for the two-gluon amplitude obtained when expanded in its color contents as in \Equation{Eq271}.
We start our analysis by collecting terms proportional to the anti-commutator of the color generators.
It can be considered the less complicated one, because it does not include terms originating from the triple-gluon vertex.
Note that it is identical with the expression for QED amplitudes of the process $e^+e^- \to \nu_\mu \bar \nu_\mu \gamma \gamma $ as described in paper \cite{Was:2004ig}.
This is possible, because the anti-commutator projects out the triple-gauge coupling.
The choice of the subtraction terms affects the final form of the results.
In case of QED we could limit ourselves to the gauge invariant parts available from the single photon (gluon) amplitude, leading to a seemingly unique result.
The case of QCD is more complex, and the choice of subtraction terms is rather motivated by aestetical considerations, for example the identification of simple gauge invariant expressions which depend simultaneously on the momenta \formula{$\pp$} and \formula{$\qq$}, leaving more lengthy expressions dependent on only one fermion momentum for further manipulations.
In fact, for the anti-commutator part, we recall the QED result, continuing the separation into parts somewhat
 further, hoping for new parts with properties which were of no interest for \cite{Was:2004ig} or even inconsistent with the structure of QED singularities.
%

\subsection{\formula{$\{\Ta,\Tb\}$}-part}
%
The part of the amplitude proportional to \formula{$\{\Ta,\Tb\}$} can be represented
as sum of twelve individually gauge invariant parts:
\begin{multline}
\Iant
=
\Iant_{1} +\Iant_{2l}+ \Iant_{2r} + \Iant_{3} +\Iant_{4l} +\Iant_{4r}
\\
+\Iant_{5lA} +\Iant_{5lB} +\Iant_{5rA} +\Iant_{5rB} +\Iant_{6l} +\Iant_{6r}
\end{multline}
where
\begin{equation}
\Iant_{1}
=
\frac{1}{2}
\,\Jsl
\bigg( \frac{\inp{\pp}{\eA}}{\inp{\pp}{\kA}} - \frac{\inp{\qq}{\eA}}{\inp{\qq}{\kA}} \bigg)
\bigg( \frac{\inp{\pp}{\eB}}{\inp{\pp}{\kB}} - \frac{\inp{\qq}{\eB}}{\inp{\qq}{\kB}} \bigg)
\label{EM7}
\end{equation}
\begin{equation}
\Iant_{2l}
=
- \frac{1}{4}
\bigg[
  \bigg( \frac{\inp{\pp}{\eA}}{\inp{\pp}{\kA}} - \frac{\inp{\qq}{\eA}}{\inp{\qq}{\kA}} \bigg)
       \frac{\esB\ksB}{\inp{\pp}{\kB}}
+ \bigg( \frac{\inp{\pp}{\eB}}{\inp{\pp}{\kB}} - \frac{\inp{\qq}{\eB}}{\inp{\qq}{\kB}} \bigg)
       \frac{\esA\ksA}{\inp{\pp}{\kA}}
\bigg]
\Jsl
\label{EM7lA}
\end{equation}
\begin{equation}
\Iant_{2r}
=
\frac{1}{4}
\,\Jsl
\bigg[
  \bigg( \frac{\inp{\pp}{\eA}}{\inp{\pp}{\kA}} - \frac{\inp{\qq}{\eA}}{\inp{\qq}{\kA}} \bigg)
       \frac{\ksB\esB}{\inp{\qq}{\kB}}
+ \bigg( \frac{\inp{\pp}{\eB}}{\inp{\pp}{\kB}} - \frac{\inp{\qq}{\eB}}{\inp{\qq}{\kB}} \bigg)
       \frac{\ksA\esA}{\inp{\qq}{\kA}}
\bigg]
\label{EM6l}
\end{equation}
\begin{equation}
\Iant_{3}
=
- \frac{1}{8}
\bigg(
  \frac{\esA\ksA}{\inp{\pp}{\kA}} \,\Jsl  \,\frac{\ksB\esB}{\inp{\qq}{\kB}}
+ \frac{\esB\ksB}{\inp{\pp}{\kB}} \,\Jsl  \,\frac{\ksA\esA}{\inp{\qq}{\kA}}
\bigg)
\label{EM5}
\end{equation}
\begin{equation}
\Iant_{4l}
=
\frac{1}{8}
\,\frac{1}{\inp{\pp}{\kA}+\inp{\pp}{\kB}-\inp{\kA}{\kB}}
\bigg( \frac{\esA\ksA\esB\ksB}{\inp{\pp}{\kA}} + \frac{\esB\ksB\esA\ksA}{\inp{\pp}{\kB}} \bigg)
\Jsl
\label{EM1}
\end{equation}
\begin{equation}
\Iant_{4r}
=
\frac{1}{8}
\,\Jsl 
\,\frac{1}{\inp{\qq}{\kA}+\inp{\qq}{\kB}-\inp{\kA}{\kB}}
\bigg( \frac{\ksB\esB\ksA\esA}{\inp{\qq}{\kA}} + \frac{\ksA\esA\ksB\esB}{\inp{\qq}{\kB}} \bigg)
\label{EM3}
\end{equation}
\begin{equation}
\Iant_{5lA}
=
\frac{1}{2}
\,\Jsl 
\,\frac{\inp{\kA}{\kB}}{\inp{\pp}{\kA}+\inp{\pp}{\kB}-\inp{\kA}{\kB}}
\bigg( \frac{\inp{\pp}{\eA}}{\inp{\pp}{\kA}} - \frac{\inp{\kB}{\eA}}{\inp{\kB}{\kA}} \bigg)
\bigg( \frac{\inp{\pp}{\eB}}{\inp{\pp}{\kB}} - \frac{\inp{\kA}{\eB}}{\inp{\kA}{\kB}} \bigg)
\label{EM2A}
\end{equation}
\begin{equation}
\Iant_{5lB}
=
- \frac{1}{2}
\,\Jsl 
\,\frac{1}{\inp{\pp}{\kA}+\inp{\pp}{\kB}-\inp{\kA}{\kB}}
\bigg( \frac{\inp{\kA}{\eB}\inp{\kB}{\eA}}{\inp{\kA}{\kB}} - \inp{\eA}{\eB} \bigg)
\label{EM2B}
\end{equation}
\begin{equation}
\Iant_{5rA}
=
\frac{1}{2}
\,\Jsl 
\,\frac{\inp{\kA}{\kB}}{\inp{\qq}{\kA}+\inp{\qq}{\kB}-\inp{\kA}{\kB}}
\bigg( \frac{\inp{\qq}{\eA}}{\inp{\qq}{\kA}} - \frac{\inp{\kB}{\eA}}{\inp{\kB}{\kA}} \bigg)
\bigg( \frac{\inp{\qq}{\eB}}{\inp{\qq}{\kB}} - \frac{\inp{\kA}{\eB}}{\inp{\kA}{\kB}} \bigg)
\label{EM4A}
\end{equation}
\begin{equation}
\Iant_{5rB}
=
- \frac{1}{2}
\,\Jsl 
\,\frac{1}{\inp{\qq}{\kA}+\inp{\qq}{\kB}-\inp{\kA}{\kB}}
\bigg( \frac{\inp{\kA}{\eB}\inp{\kB}{\eA}}{\inp{\kA}{\kB}} - \inp{\eA}{\eB} \bigg)
\label{EM4B}
\end{equation}
\begin{equation}
\Iant_{6l}
=
- \frac{1}{4}
\,\frac{\inp{\kA}{\kB}}{\inp{\pp}{\kA}+\inp{\pp}{\kB}-\inp{\kA}{\kB}}
\bigg[
  \bigg( \frac{\inp{\pp}{\eA}}{\inp{\pp}{\kA}} - \frac{\inp{\kB}{\eA}}{\inp{\kA}{\kB}} \bigg)
  \frac{\esB\ksB}{\inp{\pp}{\kB}}
+ \bigg( \frac{\inp{\pp}{\eB}}{\inp{\pp}{\kB}} - \frac{\inp{\kA}{\eB}}{\inp{\kA}{\kB}} \bigg)
  \frac{\esA\ksA}{\inp{\pp}{\kA}}
\bigg]
\Jsl
\label{EM2C}
\end{equation}
\begin{equation}
\Iant_{6r}
=
- \frac{1}{4}
\,\Jsl 
\,\frac{\inp{\kA}{\kB}}{\inp{\qq}{\kA}+\inp{\qq}{\kB}-\inp{\kA}{\kB}}
\bigg[
  \bigg( \frac{\inp{\qq}{\eA}}{\inp{\qq}{\kA}} - \frac{\inp{\kB}{\eA}}{\inp{\kA}{\kB}} \bigg)
  \frac{\ksB\esB}{\inp{\qq}{\kB}}
+ \bigg( \frac{\inp{\qq}{\eB}}{\inp{\qq}{\kB}} - \frac{\inp{\kA}{\eB}}{\inp{\kA}{\kB}} \bigg)
  \frac{\ksA\esA}{\inp{\qq}{\kA}}
\bigg]
\label{EM4C}
\end{equation}
%
This form should be compared with the one of formula~(27) in \cite{Was:2004ig}, where the amplitude is separated into seven terms \formula{$\calM_1$} to \formula{$\calM_7 $}.
Our expression \Expression{EM7} equals \formula{$\calM_7$}.
Expression \Expression{EM7lA} added to \Expression{EM6l} equals \formula{$\calM_6$}.
Expression \Expression{EM1} and \Expression{EM3} equal \formula{$\calM_1$} and \formula{$\calM_2$} respectively.
Expression \Expression{EM5}, equals \formula{$\calM_5$}.
The sum of expressions \Expression{EM2A}, \Expression{EM2B} and \Expression{EM2C} equals \formula{$\calM_3$}.
Analogously, the sum of expressions \Expression{EM4A}, \Expression{EM4B} and \Expression{EM4C} equals \formula{$\calM_4$}.

Subtraction term \Expression{a} is integrated with a positive sign into \Expression{EM7}, and with negative sign into expressions \Expression{EM4A}, \Expression{EM4B} and \Expression{EM4C}.
The analog of \Expression{a} for momentum \formula{$\pp$} is integrated into expression \Expression{EM7} with a positive sign, and into expressions \Expression{EM2A}, \Expression{EM2B}, and \Expression{EM2C} with a negative sign.

Subtraction term \Expression{b} is integrated with a positive sign into \Expression{EM7lA} and \Expression{EM6l}, and with a negative sign into expressions \Expression{EM4A}, \Expression{EM4B} and \Expression{EM4C}.
The analog of \Expression{b} for momentum \formula{$\pp$} is integrated with a positive sign into \Expression{EM7lA} and \Expression{EM6l}, and with a negative sign into expressions \Expression{EM2A}, \Expression{EM2B}, and \Expression{EM2C}.

Note that contrary to \cite{Was:2004ig} we have allowed here a new, from the point of view of fermionic spin amplitudes and Kleiss-Stirling methods \cite{Kleiss:1985yh} ugly, object \formula{$\inp{\eA}{\eB}$}.
It will appear anyway  in the commutator part discussed in the next section as a consequence of the triple gauge coupling.
We have found it  in \Expression{EM2B} and \Expression{EM4B} because of the subtraction term \Expression{c} incorporated  with the opposite sign to \Expression{EM2A} and \Expression{EM4A}.

As mentioned above, we separated the expressions \formula{$\calM_2$} and \formula{$\calM_4$} of \cite{Was:2004ig} into several new parts, partly because of the new subtraction term.
Such a separation is inappropriate for QED applications, since the new gauge invariant parts, even though compact, are more singular in the soft-photon limit than their sum, and would form an obstacle to exponentiation.
For QCD, however, this is not so much an issue because of the triple-gluon vertex;  such a separation will be useful later.
That is why we allow the structure of apparent singularities to be larger than necessary. As a consequence, the hierarchy of dominant and sub-dominant terms is richer, and potential interest in separation into more fine parts may go further.

In \cite{Was:2004ig}, the grouping of gauge invariant segments was prepared such that the terms for \formula{$\beta^0$}, \formula{$\beta^1$} and \formula{$\beta^2$} of exclusive exponentiation could be identified.
Such an approach can be productive only in combination with the discussion of a well defined calculation scheme like QED exponentiation, or at least an algorithm for phase space generation, in particular its pre-samplers and phase space Jacobians.
Therefore, the discussion of such an optimal choice of grouping would be far too pre-mature here,
 and we limit ourselves to technical properties of the amplitude.
Only when similar work for other multi-leg processes is completed as well we may return to this issue.
%

\subsection{\formula{$[\Ta,\Tb]$}-part}
For the commutator part, we introduce subtraction terms mainly in order to separate terms depending simultaneously on both momenta \formula{$\pp$} and \formula{$\qq$} into simple and intuitively clear expressions.
The part of the amplitude proportional to \formula{$[\Ta,\Tb]$} is again represented as a sum of individually gauge invariant parts and reads:
\begin{multline}
\Icom
=
\Icom_{1} +\Icom_{2l} + \Icom_{2r}+ \Icom_{3}+\Icom_{4l}+\Icom_{4r} +\Icom_{5lA} +\Icom_{5lB} +\Icom_{5rA} +\Icom_{5rB}
\\
+\Icom_{6l} +\Icom_{6r} +\Icom_{7lA} + \Icom_{7lB}+\Icom_{7rA}+ \Icom_{7rB}
\end{multline}
where
%
\begin{align}
\Icom_{1}
&=
-\frac{1}{4}
\,\Jsl
  \bigg( \frac{\inp{\pp}{\eA}}{\inp{\pp}{\kA}} - \frac{\inp{\qq}{\eA}}{\inp{\qq}{\kA}} \bigg)
  \bigg( \frac{\inp{\pp}{\eB}}{\inp{\pp}{\kB}} + \frac{\inp{\qq}{\eB}}{\inp{\qq}{\kB}} - 2  \frac{\inp{\kA}{\eB}}{\inp{\kA}{\kB}} \bigg)
\\
&+
 \frac{1}{4}
\,\Jsl
  \bigg( \frac{\inp{\pp}{\eB}}{\inp{\pp}{\kB}} - \frac{\inp{\qq}{\eB}}{\inp{\qq}{\kB}} \bigg)
  \bigg( \frac{\inp{\pp}{\eA}}{\inp{\pp}{\kA}} + \frac{\inp{\qq}{\eA}}{\inp{\qq}{\kA}} - 2  \frac{\inp{\kB}{\eA}}{\inp{\kA}{\kB}} \bigg)
\end{align}
\begin{equation}
\Icom_{2l}
=
\frac{1}{4}
\bigg[
  \bigg( \frac{\inp{\pp}{\eA}}{\inp{\pp}{\kA}} - \frac{\inp{\qq}{\eA}}{\inp{\qq}{\kA}} \bigg)  \frac{\esB\ksB}{\inp{\pp}{\kB}}
- \bigg( \frac{\inp{\pp}{\eB}}{\inp{\pp}{\kB}} - \frac{\inp{\qq}{\eB}}{\inp{\qq}{\kB}} \bigg)  \frac{\esA\ksA}{\inp{\pp}{\kA}}
\bigg]
\Jsl
\end{equation}
\begin{equation}
\Icom_{2r}
=
\frac{1}{4}
\,\Jsl
\bigg[
  \bigg( \frac{\inp{\pp}{\eA}}{\inp{\pp}{\kA}} - \frac{\inp{\qq}{\eA}}{\inp{\qq}{\kA}} \bigg)  \frac{\ksB\esB}{\inp{\qq}{\kB}}
- \bigg( \frac{\inp{\pp}{\eB}}{\inp{\pp}{\kB}} - \frac{\inp{\qq}{\eB}}{\inp{\qq}{\kB}} \bigg)  \frac{\ksA\esA}{\inp{\qq}{\kA}}
\bigg]
\end{equation}
\begin{equation}
\Icom_{3}
=
\frac{1}{8}
\bigg(
- \frac{\esA\ksA}{\inp{\pp}{\kA}} \,\Jsl  \,\frac{\ksB\esB}{\inp{\qq}{\kB}}
+ \frac{\esB\ksB}{\inp{\pp}{\kB}} \,\Jsl  \,\frac{\ksA\esA}{\inp{\qq}{\kA}}
\bigg)
\end{equation}
\begin{equation}
\Icom_{4l}
=
\frac{1}{8}
\,\frac{1}{\inp{\pp}{\kA}+\inp{\pp}{\kB}-\inp{\kA}{\kB}}
  \bigg( \frac{\esA\ksA\esB\ksB}{\inp{\pp}{\kA}} - \frac{\esB\ksB\esA\ksA}{\inp{\pp}{\kB}} \bigg)
\Jsl
\end{equation}
\begin{equation}
\Icom_{4r}
=
\frac{1}{8}
\,\Jsl 
\,\frac{1}{\inp{\qq}{\kA}+\inp{\qq}{\kB}-\inp{\kA}{\kB}}
  \bigg( \frac{\ksA\esA\ksB\esB}{\inp{\qq}{\kB}} - \frac{\ksB\esB\ksA\esA}{\inp{\qq}{\kA}} \bigg)
\end{equation}
\begin{equation}
\Icom_{5lA}
=
- \frac{1}{2}
\,\Jsl 
\,\frac{\inp{\pp}{\kA}-\inp{\pp}{\kB}}{\inp{\pp}{\kA}+\inp{\pp}{\kB}-\inp{\kA}{\kB}}
  \bigg( \frac{\inp{\pp}{\eA}}{\inp{\pp}{\kA}} - \frac{\inp{\kB}{\eA}}{\inp{\kB}{\kA}} \bigg)  \bigg(\frac{\inp{\pp}{\eB}}{\inp{\pp}{\kB}} - \frac{\inp{\kA}{\eB}}{\inp{\kA}{\kB}} \bigg)
\end{equation}
\begin{equation}
\Icom_{5lB}
=
\frac{1}{2}
\,\Jsl 
\,\frac{\inp{\pp}{\kA}-\inp{\pp}{\kB}}{\inp{\pp}{\kA}+\inp{\pp}{\kB}-\inp{\kA}{\kB}}
  \bigg( \frac{\inp{\kA}{\eB}}{\inp{\kA}{\kB}}  \frac{\inp{\kB}{\eA}}{\inp{\kA}{\kB}} - \frac{\inp{\eA}{\eB}}{\inp{\kA}{\kB}} \bigg)
\end{equation}
\begin{equation}
\Icom_{5rA}
=
- \frac{1}{2}
\,\Jsl 
\,\frac{\inp{\qq}{\kB}-\inp{\qq}{\kA}}{\inp{\qq}{\kA}+\inp{\qq}{\kB}-\inp{\kA}{\kB}}
  \bigg( \frac{\inp{\qq}{\eA}}{\inp{\qq}{\kA}} - \frac{\inp{\kB}{\eA}}{\inp{\kB}{\kA}} \bigg)  \bigg( \frac{\inp{\qq}{\eB}}{\inp{\qq}{\kB}} - \frac{\inp{\kA}{\eB}}{\inp{\kA}{\kB}} \bigg)
\end{equation}
\begin{equation}
\Icom_{5rB}
=
\frac{1}{2}
\,\Jsl 
\,\frac{\inp{\qq}{\kB}-\inp{\qq}{\kA}}{\inp{\qq}{\kA}+\inp{\qq}{\kB}-\inp{\kA}{\kB}}
  \bigg( \frac{\inp{\kA}{\eB}}{\inp{\kA}{\kB}}  \frac{\inp{\kB}{\eA}}{\inp{\kA}{\kB}} - \frac{\inp{\eA}{\eB}}{\inp{\kA}{\kB}} \bigg)
\end{equation}
\begin{equation}
\Icom_{6l}
=
\frac{1}{4}
\,\frac{\inp{\kA}{\kB}}{\inp{\pp}{\kA}+\inp{\pp}{\kB}-\inp{\kA}{\kB}}
\bigg[
+ \bigg( \frac{\inp{\pp}{\eA}}{\inp{\pp}{\kA}} - \frac{\inp{\kB}{\eA}}{\inp{\kA}{\kB}} \bigg)  \frac{\esB\ksB}{\inp{\pp}{\kB}}
- \bigg( \frac{\inp{\pp}{\eB}}{\inp{\pp}{\kB}} - \frac{\inp{\kA}{\eB}}{\inp{\kA}{\kB}} \bigg)  \frac{\esA\ksA}{\inp{\pp}{\kA}}
\bigg]
\Jsl
\end{equation}
\begin{equation}
\Icom_{6r}
=
\frac{1}{4}
\,\Jsl 
\,\frac{\inp{\kA}{\kB}}{\inp{\qq}{\kA}+\inp{\qq}{\kB}-\inp{\kA}{\kB}}
\bigg[
- \bigg( \frac{\inp{\qq}{\eA}}{\inp{\qq}{\kA}} - \frac{\inp{\kB}{\eA}}{\inp{\kA}{\kB}} \bigg)  \frac{\ksB\esB}{\inp{\qq}{\kB}}
+ \bigg( \frac{\inp{\qq}{\eB}}{\inp{\qq}{\kB}} - \frac{\inp{\kA}{\eB}}{\inp{\kA}{\kB}} \bigg)  \frac{\ksA\esA}{\inp{\qq}{\kA}}
\bigg]
\end{equation}
\begin{equation}
\Icom_{7lA}
=
\frac{1}{2}
\,\frac{1}{\inp{\pp}{\kA}+\inp{\pp}{\kB}-\inp{\kA}{\kB}}
\bigg[
- \bigg( \frac{\inp{\pp}{\eA}}{\inp{\pp}{\kA}} - \frac{\inp{\kB}{\eA}}{\inp{\kA}{\kB}} \bigg)  \esB\ksB
+ \bigg( \frac{\inp{\pp}{\eB}}{\inp{\pp}{\kB}} - \frac{\inp{\kA}{\eB}}{\inp{\kA}{\kB}} \bigg)  \esA\ksA
\bigg]
\Jsl
\end{equation}
\begin{equation}
\Icom_{7lB}
=
\frac{1}{4}
\,\frac{1}{\inp{\kA}{\kB}}\,\frac{1}{\inp{\pp}{\kA}+\inp{\pp}{\kB}-\inp{\kA}{\kB}}
\bigg( \esB\ksB\esA\ksA - \esA\ksA\esB\ksB \bigg)
\Jsl
\end{equation}
\begin{equation}
\Icom_{7rA}
=
\frac{1}{2}
\,\Jsl 
\,\frac{1}{\inp{\qq}{\kA}+\inp{\qq}{\kB}-\inp{\kA}{\kB}}
\bigg[
+ \bigg( \frac{\inp{\qq}{\eA}}{\inp{\qq}{\kA}} - \frac{\inp{\kB}{\eA}}{\inp{\kA}{\kB}} \bigg)  \ksB\esB
- \bigg( \frac{\inp{\qq}{\eB}}{\inp{\qq}{\kB}} - \frac{\inp{\kA}{\eB}}{\inp{\kA}{\kB}} \bigg)  \ksA\esA
\bigg]
\end{equation}
\begin{equation}
\Icom_{7rB}
=
- \frac{1}{4}
\,\Jsl 
\,\frac{1}{\inp{\kA}{\kB}}\,\frac{1}{\inp{\qq}{\kA}+\inp{\qq}{\kB}-\inp{\kA}{\kB}}
\bigg( \ksA\esA\ksB\esB - \ksB\esB\ksA\esA \bigg)
\end{equation}
%
This time, we do not have a good motivation for the particular form.
We mainly tried to keep it analogous to  $\Iant$ term by term.
Only the terms with a subscript starting with $7$ do not have an analogue in $\Iant$.
The subtraction terms
\begin{equation}
S^{[1,2]}_{1,\qq}
=
\Jsl
\,\frac{\inp{\kB}{\eA}}{\inp{\kA}{\kB}}
  \bigg( \frac{\inp{\pp}{\eB}}{\inp{\pp}{\kB}} -  \frac{\inp{\qq}{\eB}}{\inp{\qq}{\kB}} \bigg)
\label{aa}
\end{equation}
\begin{equation}
S^{[1,2]}_{2,\qq}
=
- \Jsl
\bigg(
  \frac{\ksB\esB}{\inp{\qq}{\kB}} \; \frac{\inp{\qq}{\eA}}{\inp{\qq}{\kA}}
- \frac{\ksA\esA}{\inp{\qq}{\kA}} \; \frac{\inp{\qq}{\eB}}{\inp{\qq}{\kB}}
\bigg)
\label{bb}
\end{equation}

\begin{equation}
S^{[1,2]}_{3,\qq}
=
\frac{1}{2}
\,\Jsl 
\frac{1}{\inp{\qq}{\kA}+\inp{\qq}{\kB}- \inp{\kA}{\kB}}
\Bigl(
- \frac{2\inp{\qq}{\eB}\inp{\kB}{\eA}}{\inp{\qq}{\kB}}
+ \frac{2\inp{\qq}{\eA}\inp{\kA}{\eB}}{\inp{\qq}{\kA}}
\Bigr)
\label{cc}
~
\end{equation}
are quite analogous to the ones used in the previous subsection and we will not elaborate on them further.
The \formula{$S^{[1,2]}_{1,\pp}$}, \formula{$S^{[1,2]}_{2,\pp}$} and \formula{$S^{[1,2]}_{3,\pp}$} can be obtained analogously.

\section{Picture of consecutive gluon emission\label{SecSingProd}}
Encouraged by the rich substructure of gauge invariant parts we found in our expressions, we search for other possible organizations of terms.
In particular, we try to incorporate an ordering of the gluons, which has proved its usefulness in a broad spectrum of applications and calculations concerning QCD.

\subsection{Color-ordered amplitudes\label{TAB}}
Instead of an expansion in the commutator and the anti-commutator of color generators, which is useful for the comparison with existing QED results, let us simply use generator products to express
%
\begin{equation}
\Mab
=
\frac{1}{2}
\,\bar{v}(\pp)
\Big(
\Ta\Tb \IAB +\Tb\Ta \IBA
\Big)
u(\qq)
~.
\end{equation}
%
The so-called color-ordered amplitudes are obtained from the coefficients \formula{$\IAB$} and \formula{$\IBA$} by including the spinors $\bar{v}(\pp)$ and $u(\qq)$.
Thanks to this ordering the expressions shorten significantly.
%
%
For the $\Ta\Tb$-part, we find
\begin{align}
\IAB
&=
\bigg(
 \frac{\inp{\pp}{\eA}}{\inp{\pp}{\kA}}
-\frac{\inp{\kB}{\eA}}{\inp{\kB}{\kA}}
-\frac{\esA\ksA}{2\inp{\pp}{\kA}}
\bigg)
\Jsl
\bigg(
 \frac{\ksB\esB}{2\inp{\qq}{\kB}}
+\frac{\inp{\kA}{\eB}}{\inp{\kA}{\kB}}
-\frac{\inp{\qq}{\eB}}{\inp{\qq}{\kB}}
\bigg)
\label{abstart}
\\
&+
\frac{\inp{\pp}{\kB}}{\inp{\pp}{\kA}+\inp{\pp}{\kB}-\inp{\kA}{\kB}}
\,\bigg(\frac{\inp{\pp}{\eA}}{\inp{\pp}{\kA}}-\frac{\inp{\kB}{\eA}}{\inp{\kB}{\kA}}
        -\frac{\esA\ksA}{2\inp{\pp}{\kA}}\bigg)
  \bigg(\frac{\inp{\pp}{\eB}}{\inp{\pp}{\kB}}-\frac{\inp{\kA}{\eB}}{\inp{\kA}{\kB}}
        -\frac{\esB\ksB}{2\inp{\pp}{\kB}}\bigg)
\Jsl
\label{ab-2}
\\
&+
\Jsl 
\,\frac{\inp{\qq}{\kA}}{\inp{\qq}{\kA}+\inp{\qq}{\kB}-\inp{\kA}{\kB}}
\,\bigg(\frac{\inp{\qq}{\eA}}{\inp{\qq}{\kA}}-\frac{\inp{\kB}{\eA}}{\inp{\kB}{\kA}}
        -\frac{\ksA\esA}{2\inp{\qq}{\kA}}\bigg)
  \bigg(\frac{\inp{\qq}{\eB}}{\inp{\qq}{\kB}}-\frac{\inp{\kA}{\eB}}{\inp{\kA}{\kB}}
        -\frac{\ksB\esB}{2\inp{\qq}{\kB}}\bigg)
\label{ab-3}
\\
&+
\Jsl 
\bigg(
1
 -\frac{\inp{\pp}{\kB}}{\inp{\pp}{\kA}+\inp{\pp}{\kB}-\inp{\kA}{\kB}}
-\frac{\inp{\qq}{\kA}}{\inp{\qq}{\kA}+\inp{\qq}{\kB}-\inp{\kA}{\kB}}
\bigg)
\bigg( \frac{\inp{\kA}{\eB}}{\inp{\kA}{\kB}}\,\frac{\inp{\kB}{\eA}}{\inp{\kA}{\kB}}
      -\frac{\inp{\eA}{\eB}}{\inp{\kA}{\kB}}\bigg)
\label{abrun}
\\
&-
\frac{1}{4}
\,\frac{1}{\inp{\pp}{\kA}+\inp{\pp}{\kB}-\inp{\kA}{\kB}}
\bigg(
  \frac{ \esA\ksA\esB\ksB
         -\esB\ksB\esA\ksA }{\inp{\kA}{\kB}}
\bigg)
\Jsl
\label{ab-4}
\\
&-
\frac{1}{4}
\,\Jsl 
\,\frac{1}{\inp{\qq}{\kA}+\inp{\qq}{\kB}-\inp{\kA}{\kB}}
\bigg(
  \frac{ \ksA\esA\ksB\esB
         -\ksB\esB\ksA\esA }{\inp{\kA}{\kB}}
\bigg)
\label{abend}
~.
\end{align}
%
The part proportional to \formula{$\Tb\Ta$} is obtained by a permutation of the momenta and polarization vectors of the gluons, and reads
%
\begin{align}
\IBA
&=
\bigg(
 \frac{\inp{\pp}{\eB}}{\inp{\pp}{\kB}}
-\frac{\inp{\kA}{\eB}}{\inp{\kA}{\kB}}
-\frac{\esB\ksB}{2\inp{\pp}{\kB}}
\bigg)
\Jsl
\bigg(
 \frac{\ksA\esA}{2\inp{\qq}{\kA}}
+\frac{\inp{\kB}{\eA}}{\inp{\kB}{\kA}}
-\frac{\inp{\qq}{\eA}}{\inp{\qq}{\kA}}
\bigg)
\label{bastart}
\\
&+
\frac{\inp{\pp}{\kA}}{\inp{\pp}{\kB}+\inp{\pp}{\kA}-\inp{\kB}{\kA}}
\,\bigg(\frac{\inp{\pp}{\eB}}{\inp{\pp}{\kB}}-\frac{\inp{\kA}{\eB}}{\inp{\kA}{\kB}}
        -\frac{\esB\ksB}{2\inp{\pp}{\kB}}\bigg)
  \bigg(\frac{\inp{\pp}{\eA}}{\inp{\pp}{\kA}}-\frac{\inp{\kB}{\eA}}{\inp{\kB}{\kA}}
        -\frac{\esA\ksA}{2\inp{\pp}{\kA}}\bigg)
\Jsl
\label{ba-2}
\\
&+
\Jsl 
\,\frac{\inp{\qq}{\kB}}{\inp{\qq}{\kB}+\inp{\qq}{\kA}-\inp{\kB}{\kA}}
\,\bigg(\frac{\inp{\qq}{\eB}}{\inp{\qq}{\kB}}-\frac{\inp{\kA}{\eB}}{\inp{\kA}{\kB}}
        -\frac{\ksB\esB}{2\inp{\qq}{\kB}}\bigg)
  \bigg(\frac{\inp{\qq}{\eA}}{\inp{\qq}{\kA}}-\frac{\inp{\kB}{\eA}}{\inp{\kB}{\kA}}
        -\frac{\ksA\esA}{2\inp{\qq}{\kA}}\bigg)
\label{ba-3}
\\
&+
\Jsl 
\bigg(
1
 -\frac{\inp{\pp}{\kA}}{\inp{\pp}{\kB}+\inp{\pp}{\kA}-\inp{\kB}{\kA}}
-\frac{\inp{\qq}{\kB}}{\inp{\qq}{\kB}+\inp{\qq}{\kA}-\inp{\kB}{\kA}}
\bigg)
\bigg( \frac{\inp{\kB}{\eA}}{\inp{\kB}{\kA}}\,\frac{\inp{\kA}{\eB}}{\inp{\kB}{\kA}}
      -\frac{\inp{\eB}{\eA}}{\inp{\kB}{\kA}}\bigg)
\label{barun}
\\
&-
\frac{1}{4}
\,\frac{1}{\inp{\pp}{\kB}+\inp{\pp}{\kA}-\inp{\kB}{\kA}}
\bigg(
  \frac{ \esB\ksB\esA\ksA
         -\esA\ksA\esB\ksB }{\inp{\kB}{\kA}}
\bigg)
\Jsl
\label{ba-4}
\\
&-
\frac{1}{4}
\,\Jsl 
\,\frac{1}{\inp{\qq}{\kB}+\inp{\qq}{\kA}-\inp{\kB}{\kA}}
\bigg(
  \frac{ \ksB\esB\ksA\esA
         -\ksA\esA\ksB\esB }{\inp{\kB}{\kA}}
\bigg)
\label{baend}
~.
\end{align}
%
Each line of the above expressions is individually gauge invariant.

%
\subsection{Mixed representation\label{MixedR}}
%
As we will see now, the expression for the amplitude obtained with a mixed, over-defined, basis will not only be shorter than the previously discussed ones, but will also lend itself better to analyze the properties of the amplitude.
Thanks to such a representation not only the exact results, but also expressions dominant in some regions of the phase space (obtained simply by truncation) will be more compact.
The decomposition of the amplitude looks as follows
%
\begin{equation}
\Mab
=
\frac{1}{2}
\,\bar{v}(\pp)\Big(
\Ta\Tb\ImixAB + \Tb\Ta\ImixBA + [\Ta,\Tb]\Imixcom + \{\Ta,\Tb\}\Imixant
\Big)u(\qq)
~,
\label{Mab}
\end{equation}
%
with the coefficients given by
%
\begin{align}
\ImixAB
&=
\bigg(
 \frac{\inp{\pp}{\eA}}{\inp{\pp}{\kA}}
-\frac{\inp{\kB}{\eA}}{\inp{\kB}{\kA}}
-\frac{\esA\ksA}{2\inp{\pp}{\kA}}
\bigg)
\Jsl
\bigg(
 \frac{\ksB\esB}{2\inp{\qq}{\kB}}
+\frac{\inp{\kA}{\eB}}{\inp{\kA}{\kB}}
-\frac{\inp{\qq}{\eB}}{\inp{\qq}{\kB}}
\bigg)
\nonumber\\
&+
\frac{\inp{\pp}{\kB}}{\inp{\pp}{\kA}+\inp{\pp}{\kB}-\inp{\kA}{\kB}}
\,\bigg(\frac{\inp{\pp}{\eA}}{\inp{\pp}{\kA}}-\frac{\inp{\kB}{\eA}}{\inp{\kB}{\kA}}
        -\frac{\esA\ksA}{2\inp{\pp}{\kA}}\bigg)
  \bigg(\frac{\inp{\pp}{\eB}}{\inp{\pp}{\kB}}-\frac{\inp{\kA}{\eB}}{\inp{\kA}{\kB}}
        -\frac{\esB\ksB}{2\inp{\pp}{\kB}}\bigg)
\Jsl
\phantom{xx}
\label{ImixAB}\\
&+
\Jsl 
\,\frac{\inp{\qq}{\kA}}{\inp{\qq}{\kA}+\inp{\qq}{\kB}-\inp{\kA}{\kB}}
\,\bigg(\frac{\inp{\qq}{\eA}}{\inp{\qq}{\kA}}-\frac{\inp{\kB}{\eA}}{\inp{\kB}{\kA}}
        -\frac{\ksA\esA}{2\inp{\qq}{\kA}}\bigg)
  \bigg(\frac{\inp{\qq}{\eB}}{\inp{\qq}{\kB}}-\frac{\inp{\kA}{\eB}}{\inp{\kA}{\kB}}
        -\frac{\ksB\esB}{2\inp{\qq}{\kB}}\bigg)
\nonumber
\end{align}
%
\begin{align}
\ImixBA
&=
\bigg(
 \frac{\inp{\pp}{\eB}}{\inp{\pp}{\kB}}
-\frac{\inp{\kA}{\eB}}{\inp{\kA}{\kB}}
-\frac{\esB\ksB}{2\inp{\pp}{\kB}}
\bigg)
\Jsl
\bigg(
 \frac{\ksA\esA}{2\inp{\qq}{\kA}}
+\frac{\inp{\kB}{\eA}}{\inp{\kB}{\kA}}
-\frac{\inp{\qq}{\eA}}{\inp{\qq}{\kA}}
\bigg)
\nonumber\\
&+
\frac{\inp{\pp}{\kA}}{\inp{\pp}{\kB}+\inp{\pp}{\kA}-\inp{\kB}{\kA}}
\,\bigg(\frac{\inp{\pp}{\eB}}{\inp{\pp}{\kB}}-\frac{\inp{\kA}{\eB}}{\inp{\kA}{\kB}}
        -\frac{\esB\ksB}{2\inp{\pp}{\kB}}\bigg)
  \bigg(\frac{\inp{\pp}{\eA}}{\inp{\pp}{\kA}}-\frac{\inp{\kB}{\eA}}{\inp{\kB}{\kA}}
        -\frac{\esA\ksA}{2\inp{\pp}{\kA}}\bigg)
\Jsl
\phantom{xx}
\label{ImixBA}\\
&+
\Jsl 
\,\frac{\inp{\qq}{\kB}}{\inp{\qq}{\kB}+\inp{\qq}{\kA}-\inp{\kB}{\kA}}
\,\bigg(\frac{\inp{\qq}{\eB}}{\inp{\qq}{\kB}}-\frac{\inp{\kA}{\eB}}{\inp{\kA}{\kB}}
        -\frac{\ksB\esB}{2\inp{\qq}{\kB}}\bigg)
  \bigg(\frac{\inp{\qq}{\eA}}{\inp{\qq}{\kA}}-\frac{\inp{\kB}{\eA}}{\inp{\kB}{\kA}}
        -\frac{\ksA\esA}{2\inp{\qq}{\kA}}\bigg)
\nonumber
\end{align}
%
\begin{align}
\Imixcom
&=
\frac{1}{2}
\,\Jsl 
\bigg(
  \frac{\inp{\pp}{\kA}-\inp{\pp}{\kB}}{\inp{\pp}{\kA}+\inp{\pp}{\kB}-\inp{\kA}{\kB}}
+ \frac{\inp{\qq}{\kB}-\inp{\qq}{\kA}}{\inp{\qq}{\kA}+\inp{\qq}{\kB}-\inp{\kA}{\kB}}
\bigg)
\bigg( \frac{\inp{\kA}{\eB}}{\inp{\kA}{\kB}}\,\frac{\inp{\kB}{\eA}}{\inp{\kA}{\kB}}
      -\frac{\inp{\eA}{\eB}}{\inp{\kA}{\kB}}\bigg)
\nonumber\\
&-
\frac{1}{4}
\,\frac{1}{\inp{\pp}{\kA}+\inp{\pp}{\kB}-\inp{\kA}{\kB}}
\bigg(
  \frac{ \esA\ksA\esB\ksB -\esB\ksB\esA\ksA }{\inp{\kA}{\kB}}
\bigg)
\Jsl
\label{Imixcom}\\
&-
\frac{1}{4}
\,\Jsl 
\,\frac{1}{\inp{\qq}{\kA}+\inp{\qq}{\kB}-\inp{\kA}{\kB}}
\bigg(
  \frac{ \ksA\esA\ksB\esB -\ksB\esB\ksA\esA }{\inp{\kA}{\kB}}
\bigg)
\nonumber
\end{align}
%
\begin{equation}
\Imixant
=
-\frac{1}{2}
\,\Jsl 
\bigg(
  \frac{1}{\inp{\pp}{\kA}+\inp{\pp}{\kB}-\inp{\kA}{\kB}}
+ \frac{1}{\inp{\qq}{\kA}+\inp{\qq}{\kB}-\inp{\kA}{\kB}}
\bigg)
\bigg(
  \frac{\inp{\kA}{\eB}\;\inp{\kB}{\eA}}{\inp{\kA}{\kB}} -\inp{\eA}{\eB}
\bigg)
\label{Imixant}
~.
\end{equation}
%

The choice for the above representation is not unique, and at this point
it seems to be based mainly on aesthetic grounds.
Again each line of the above expressions is individually gauge invariant.
In order to justify our choice a bit further, we change our strategy in the construction of the expressions.
So far, the only manipulations we performed consisted of the reorganization of terms, and the introduction of our so-called subtraction terms.
All these terms were written explicitly in terms of momenta and polarization vectors.
Now, we will leave this path, and start to introduce new objects useful to
compactify the  expression even further.
The objects are the momentum of the virtual gluon
%
\begin{equation}
\kAB^{\mu}
=
\kA^{\mu} + \kB^{\mu}
~,
\end{equation}
%
and the four-vectors
%
\begin{equation}
\eAB^{\mu}
=
\frac{\kA^{\mu}-\kB^{\mu}}{2\,\inp{\kA}{\kB}}
\bigg(
 \frac{\inp{\eA}{\kB}\;\inp{\eB}{\kA}}{\inp{\kA}{\kB}}-\inp{\eA}{\eB}
\bigg)
\quad,\quad
\fAB^{\mu}
=
\frac{\kA^{\mu}-\kB^{\mu}}{2\,\inp{\kA}{\kB}}
\;
\frac{\frac{\imag}{4}\,\Tr(\gamma^{5}\esA\ksA\esB\ksB)}{\inp{\kA}{\kB}}
~,
\label{Exp1347}
\end{equation}
%
which represent the effective polarizations of the virtual gluon piled together with its propagator.%
\footnote{Proper units of energy will appear only after including the appropriate factor from the phase space Jacobian (this is usually done, only after some assumption is made on phase space regions to be considered).
That is why wrong units for this polarization vector are not as disastrous as they may seem.
Note also, that $\eAB$ is parallel to $\fAB$ and has a significant component along the $\kAB$-direction.
This is another reason to be cautious with their physical interpretation.
}
Notice, that the expressions are gauge-invariant.
We immediately see that the 
first line of \Equation{Imixcom} is equal to
%
\begin{equation}
\Jsl 
\bigg(
  \frac{\inp{\pp}{\eAB}}{\inp{\pp}{\kAB}-\inp{\kA}{\kB}}
- \frac{\inp{\qq}{\eAB}}{\inp{\qq}{\kAB}-\inp{\kA}{\kB}}
\bigg)
~.
\end{equation}
%
Furthermore, we will prove in the next section that
%
\begin{equation}
\frac{\esA\ksA\esB\ksB-\esB\ksB\esA\ksA}{2\inp{\kA}{\kB}}
=
(\esAB+\imag\gamma^{5}\fsAB)\ksAB
~,
\label{Eq1378}
\end{equation}
%
so that we can write
%
\begin{align}
\Imixcom
&=
\Jsl 
\bigg(
  \frac{\inp{\pp}{\eAB}}{\inp{\pp}{\kAB}-\inp{\kA}{\kB}}
- \frac{\inp{\qq}{\eAB}}{\inp{\qq}{\kAB}-\inp{\kA}{\kB}}
\bigg)
\nonumber\\
&-
\frac{1}{2}
\,\frac{1}{\inp{\pp}{\kAB}-\inp{\kA}{\kB}}
\big(\esAB+\imag\gamma^{5}\fsAB\big)\ksAB
\Jsl
\label{Imixcom1}\\
&+
\frac{1}{2}
\,\Jsl 
\,\frac{1}{\inp{\qq}{\kAB}-\inp{\kA}{\kB}}
\,\ksAB\big(\esAB - \imag\gamma^{5}\fsAB\big)
\nonumber
~.
\end{align}
%
Notice the similarity of this expression with \Equation{firstO} for the single-gluon emission.
In the next section, we will show that in the limit when \formula{$\inp{\kA}{\kB}$} becomes zero, whether it be a soft or a collinear limit, the quantities
\begin{equation}
\bigg(\frac{\inp{\eA}{\kB}\;\inp{\eB}{\kA}}{\inp{\kA}{\kB}}-\inp{\eA}{\eB}\bigg)
\quad\textrm{and}\quad
\frac{\frac{\imag}{4}\,\Tr(\gamma^{5}\esA\ksA\esB\ksB)}{\inp{\kA}{\kB}}
\label{Eq1426}
\end{equation}
%
stay both finite.
In particular, when the polarization vectors \formula{$\eA$} and \formula{$\eB$} become parallel, $\fAB$, absent anyway in the first line of \Expression{Imixcom1}, becomes zero, indicating that it contributes negligibly compared to \formula{$\eAB$}, and making the \Expression{Imixcom1} even more similar to \Expression{firstO}.
Also,
%
\begin{equation} 
\inp{\bigg(
 \frac{\pp}{\inp{\pp}{\kA}+\inp{\pp}{\kB}-\inp{\kA}{\kB}}
-\frac{\qq}{\inp{\qq}{\kA}+\inp{\qq}{\kB}-\inp{\kA}{\kB}}\bigg)}
{\frac{\kA-\kB}{\inp{\kA}{\kB}}}
\label{Eq1448}
\end{equation} 
%
remains finite when $\kA$ and $\kB$ become collinear. 
This property of the expression present in the first line of \Expression{Imixcom}
 is a technical manifestation of the absence 
of longitudinal gluons, but this time for the virtual gluon. 
Again, the similarity to \Expression{firstO} manifests itself. There, the cancellation is exact, and is a consequence of gauge invariance in combination with the fact that the gluon is on mass-shell.

As mentioned, the proof of \Equation{Eq1378} will be given in the next section.
Here, we just want to mention that
%
\begin{equation}
\imag\gamma^{5}\fsAB\ksAB
=
\frac{\esA\ksA\esB\ksB-\esB\ksB\esA\ksA}{4\inp{\kA}{\kB}}
+
\frac{\ksA\ksB\esB\esA\ksA\ksB-\ksB\ksA\esA\esB\ksB\ksA}{8\inp{\kA}{\kB}\;\inp{\kA}{\kB}}
~.
\end{equation}
%

\subsection{Towards an interpretation at the exact amplitude level}
%
The expression for the amplitude can be compactified further with the help of the four-vectors
%
\begin{equation}
\fA^{\mu}
=
\eA^{\mu} - \frac{\inp{\eA}{\kB}}{\inp{\kA}{\kB}}\,\kA^{\mu}
\quad\textrm{and}\quad
\fB^{\mu}
=
\eB^{\mu} - \frac{\inp{\eB}{\kA}}{\inp{\kA}{\kB}}\,\kB^{\mu}
~.
\end{equation}
%
These manifestly gauge-invariant objects satisfy
%
\begin{equation}
 \inp{\fA}{\kA}
=\inp{\fA}{\kB}
=\inp{\fB}{\kA}
=\inp{\fB}{\kB}
=0
~,
\label{Exp60}
\end{equation}
%
and can be understood as a particular choice of gauge for $\eA$, $\eB$.
Alternatively the terms \formula{$\kA^{\mu}\,\inp{\eA}{\kB}/\inp{\kA}{\kB}$} and \formula{$\kB^{\mu}\,\inp{\eB}{\kA}/\inp{\kA}{\kB}$} can be understood as defining subtraction terms.
The coefficients of the amplitude \Expression{Mab} can now be written as follows:
%
\begin{align}
\ImixAB
&=
\bigg(
 \frac{\inp{\pp}{\fA}}{\inp{\pp}{\kA}}
-\frac{\fsA\ksA}{2\inp{\pp}{\kA}}
\bigg)
\Jsl
\bigg(
 \frac{\ksB\fsB}{2\inp{\qq}{\kB}}
-\frac{\inp{\qq}{\fB}}{\inp{\qq}{\kB}}
\bigg)
\nonumber\\
&+
\frac{\inp{\pp}{\kB}}{\inp{\pp}{\kA}+\inp{\pp}{\kB}-\inp{\kA}{\kB}}
\,\bigg(\frac{\inp{\pp}{\fA}}{\inp{\pp}{\kA}}
       -\frac{\fsA\ksA}{2\inp{\pp}{\kA}}\bigg)
  \bigg(\frac{\inp{\pp}{\fB}}{\inp{\pp}{\kB}}
       -\frac{\fsB\ksB}{2\inp{\pp}{\kB}}\bigg)
\Jsl
\\
&+
\Jsl 
\,\frac{\inp{\qq}{\kA}}{\inp{\qq}{\kA}+\inp{\qq}{\kB}-\inp{\kA}{\kB}}
\,\bigg(\frac{\inp{\qq}{\fA}}{\inp{\qq}{\kA}}
       -\frac{\ksA\fsA}{2\inp{\qq}{\kA}}\bigg)
  \bigg(\frac{\inp{\qq}{\fB}}{\inp{\qq}{\kB}}
       -\frac{\ksB\fsB}{2\inp{\qq}{\kB}}\bigg)
\nonumber
\end{align}
%
\begin{align}
\ImixBA
&=
\bigg(
 \frac{\inp{\pp}{\fB}}{\inp{\pp}{\kB}}
-\frac{\fsB\ksB}{2\inp{\pp}{\kB}}
\bigg)
\Jsl
\bigg(
 \frac{\ksA\fsA}{2\inp{\qq}{\kA}}
-\frac{\inp{\qq}{\fA}}{\inp{\qq}{\kA}}
\bigg)
\nonumber\\
&+
\frac{\inp{\pp}{\kA}}{\inp{\pp}{\kB}+\inp{\pp}{\kA}-\inp{\kB}{\kA}}
\,\bigg(\frac{\inp{\pp}{\fB}}{\inp{\pp}{\kB}}
       -\frac{\fsB\ksB}{2\inp{\pp}{\kB}}\bigg)
  \bigg(\frac{\inp{\pp}{\fA}}{\inp{\pp}{\kA}}
       -\frac{\fsA\ksA}{2\inp{\pp}{\kA}}\bigg)
\Jsl
\\
&+
\Jsl 
\,\frac{\inp{\qq}{\kB}}{\inp{\qq}{\kB}+\inp{\qq}{\kA}-\inp{\kB}{\kA}}
\,\bigg(\frac{\inp{\qq}{\fB}}{\inp{\qq}{\kB}}
       -\frac{\ksB\fsB}{2\inp{\qq}{\kB}}\bigg)
  \bigg(\frac{\inp{\qq}{\fA}}{\inp{\qq}{\kA}}
       -\frac{\ksA\fsA}{2\inp{\qq}{\kA}}\bigg)
\nonumber
\end{align}
%
\begin{align}
\Imixcom
&=
\frac{1}{2}
\,\Jsl 
\bigg(
  \frac{\inp{\pp}{\kA}-\inp{\pp}{\kB}}{\inp{\pp}{\kA}+\inp{\pp}{\kB}-\inp{\kA}{\kB}}
+ \frac{\inp{\qq}{\kB}-\inp{\qq}{\kA}}{\inp{\qq}{\kA}+\inp{\qq}{\kB}-\inp{\kA}{\kB}}
\bigg)
\bigg( -\frac{\inp{\fA}{\fB}}{\inp{\kA}{\kB}}\bigg)
\nonumber\\
&-
\frac{1}{4}
\,\frac{1}{\inp{\pp}{\kA}+\inp{\pp}{\kB}-\inp{\kA}{\kB}}
\bigg(
  \frac{ \fsA\ksA\fsB\ksB -\fsB\ksB\fsA\ksA }{\inp{\kA}{\kB}}
\bigg)
\Jsl
\\
&-
\frac{1}{4}
\,\Jsl 
\,\frac{1}{\inp{\qq}{\kA}+\inp{\qq}{\kB}-\inp{\kA}{\kB}}
\bigg(
  \frac{ \ksA\fsA\ksB\fsB -\ksB\fsB\ksA\fsA }{\inp{\kA}{\kB}}
\bigg)
\nonumber
\end{align}
%
\begin{equation}
\Imixant
=
-\frac{1}{2}
\,\Jsl 
\bigg(
  \frac{1}{\inp{\pp}{\kA}+\inp{\pp}{\kB}-\inp{\kA}{\kB}}
+ \frac{1}{\inp{\qq}{\kA}+\inp{\qq}{\kB}-\inp{\kA}{\kB}}
\bigg)
\big(
  -\inp{\fA}{\fB}
\big)
\end{equation}
%
The four-vectors $\fA$, $\fB$ appear to be particularly useful for the proofs promised in the previous section.
Let us note first that in terms of these four-vectors the ``polarization vectors'' of the virtual gluon defined in \Expression{Exp1347} are given by
%
\begin{equation}
\eAB^{\mu}
=
\frac{\kA^{\mu}-\kB^{\mu}}{2\,\inp{\kA}{\kB}}
\big(
 -\inp{\fA}{\fB}
\big)
\quad,\quad
\fAB^{\mu}
=
\frac{\kA^{\mu}-\kB^{\mu}}{2\,\inp{\kA}{\kB}}
\;
\frac{\frac{\imag}{4}\,\Tr(\gamma^{5}\fsA\ksA\fsB\ksB)}{\inp{\kA}{\kB}}
~,
\end{equation}
%
and that
%
\begin{equation}
\esA\ksA\esB\ksB-\esB\ksB\esA\ksA
=
\fsA\ksA\fsB\ksB-\fsB\ksB\fsA\ksA
~.
\end{equation}
%
To proof \Equation{Eq1378}, we will first show that the formula holds in a particular Lorentz-frame.
Since the expressions are manifestly Lorentz invariant, that is sufficient for the formula to hold in general case as well.
Let us choose a rest-frame of the gluon-pair in which \formula{$\kA$} and \formula{$\kB$} are back-to-back  and \formula{$\fA$} points along the $x$-axis:
\newcommand{\kk}{k}
\newcommand{\Cos}{c}
\newcommand{\Sin}{s}
\begin{equation}
\kA^{\mu}=\kk(1,0,0,1)
\quad,\quad
\kB^{\mu}=\kk(1,0,0,-1)
\quad,\quad
\fA^{\mu}=(0,1,0,0)
\quad,\quad
\fB^{\mu}=(0,\Cos,\Sin,0)
~,
\end{equation}
%
with
%
\begin{equation}
\kk = \sqrt{\inp{\kA}{\kB}/2}
\quad\textrm{and}\quad
\Cos^2+\Sin^2 = 1
~.
\end{equation}
%
In our frame, we have
%
\begin{equation}
\fsA\ksA\fsB\ksB-\fsB\ksB\fsA\ksA
= -(\ksA\ksB-\ksB\ksA)\Cos - 4\kk^2\gamma^{1}\gamma^{2}\Sin
~.
\end{equation}
%
Observing that \formula{$\gamma^{1}\gamma^{2}=\imag\gamma^{5}\gamma^{0}\gamma^{3}$}, \formula{$\ksA+\ksB=2\kk\gamma^0$} and that \formula{$\ksA-\ksB=-2\kk\gamma^{3}$}, we rewrite the above expression as
\footnote{%
We use the definition \formula{$\gamma^{5}=\imag\gamma^{0}\gamma^{1}\gamma^{2}\gamma^{3}$}.
}%
%
\begin{equation}
\fsA\ksA\fsB\ksB-\fsB\ksB\fsA\ksA
= -(\ksA\ksB-\ksB\ksA)(\Cos +\imag\gamma^{5}\Sin)
~.
\end{equation}
%
On the other hand, we have
%
\begin{equation}
(2\inp{\kA}{\kB})\esAB\ksAB
=
(\ksA\ksB-\ksB\ksA)(-\inp{\fA}{\fB})
=
-(\ksA\ksB-\ksB\ksA)\Cos
~,
\end{equation}
%
and
%
\begin{equation}
(2\inp{\kA}{\kB})\fsAB\ksAB
=
(\ksA\ksB-\ksB\ksA)
\frac{\frac{\imag}{4}\,\Tr(\gamma^{5}\fsA\ksA\fsB\ksB)}{\inp{\kA}{\kB}}
=
-(\ksA\ksB-\ksB\ksA)\Sin
~.
\end{equation}
%
Now, it is straightforward to complete the proof of \Expression{Eq1378} by elimination of the  (frame dependent) expressions on the r.h.s.\ of the above equations.

For the investigation of the behavior of \Expression{Eq1426} in the limit in which \formula{$\kA$} and \formula{$\kB$} become collinear, we should not use a Lorentz frame in which these momenta are forced to be back-to-back.
We keep the three-vector of \formula{$\kA$} along the $z$-axis but give the three-vector of \formula{$\kB$} a component along the $y$-axis:
%
\begin{equation}
\kA^{\mu}
=
\kA^{0}(1,0,0,1)
\quad,\quad
\kB^{\mu}
=
\kB^{0}(1,0,y,z)
\end{equation}
%
with
%
\begin{equation}
y^2 = 1-z^2 = (1-z)(1+z)
~.
\end{equation}
%
The collinear limit corresponds to the limit \formula{$z\to1$}.
According to \Expression{Exp60}, the four-vectors \formula{$\fA$} and \formula{$\fB$} must have the form
\newcommand{\fAone}{x_1}
\newcommand{\fBone}{x_2}
%
\begin{equation}
\fA^{\mu}
=
\bigg(\sqrt{\frac{1+z}{1-z}}\,\Sin_1,\Cos_1,\Sin_1,\sqrt{\frac{1+z}{1-z}}\,\Sin_1\bigg)
\quad,\quad
\fB^{\mu}
=
\bigg(\sqrt{\frac{1+z}{1-z}}\,\Sin_2,\Cos_2,\Sin_2,\sqrt{\frac{1+z}{1-z}}\,\Sin_2\bigg)
~,
\end{equation}
%
with the only further restriction that
%
\begin{equation}
\Cos_1^2+\Sin_1^2 = -\inp{\eA}{\eA}
\quad,\quad
\Cos_2^2+\Sin_2^2 = -\inp{\eB}{\eB}
~.
\end{equation}
%
Explicit calculation gives
%
\begin{equation}
\bigg(\frac{\inp{\eA}{\kB}\;\inp{\eB}{\kA}}{\inp{\kA}{\kB}}-\inp{\eA}{\eB}\bigg)
=
-\inp{\fA}{\fB}
=
\Cos_1\Cos_2 + \Sin_1\Sin_2
~,
\end{equation}
%
and
%
\begin{equation}
\frac{\frac{\imag}{4}\,\Tr(\gamma^{5}\esA\ksA\esB\ksB)}{\inp{\kA}{\kB}}
=
\frac{\frac{\imag}{4}\,\Tr(\gamma^{5}\fsA\ksA\fsB\ksB)}{\inp{\kA}{\kB}}
=
\Cos_1\Sin_2 - \Cos_2\Sin_1
~,
\end{equation}
%
and we see that both quantities are independent of $z$.

\subsection{Properties of the amplitudes\label{cztery-cztery}}
%
Before we discuss the properties of the expressions under certain kinematical constraints, \eg\ orderings, let us analyze here the building blocks which appeared already now, at the exact amplitude level.
For that purpose we will exploit the form of the amplitude as given in \Section{TAB} and \Section{MixedR}.
In principle, such organization remains valid independently of whether there are gluons or fermions in the initial state.
Nonetheless, we will progressively start to focus our attention on the case where the incoming states are fermions and carry dominant momenta; \formula{$\pp$} and \formula{$\qq$} respectively.

In Section~13.1 of \cite{Was:1994kg} it was shown in a pedagogical manner, that for example in case \formula{$\kA$} becomes collinear with \formula{$\pp$} the factor
%
\begin{equation}
\bigg(
 \frac{\inp{\pp}{\eA}}{\inp{\pp}{\kA}}
-\frac{\esA\ksA}{2\inp{\pp}{\kA}}
\bigg)
\label{LL}
\end{equation}
%
gives, after phase space integration over the appropriate region, the Altarelli-Parisi kernel.
It can thus be understood as its precursor at the spin amplitude level.
Like this, of course, such an expression makes no sense, since it is gauge-dependent.
Only a gauge invariant-object like
%
\begin{equation}
\bigg(
 \frac{\inp{\pp}{\eA}}{\inp{\pp}{\kA}}
-\frac{\inp{\kB}{\eA}}{\inp{\kB}{\kA}}
-\frac{\esA\ksA}{2\inp{\pp}{\kA}}
\bigg)
\label{dipol}
\end{equation}
%
can be interpreted unambiguously.
It can be understood as a precursor at the spin amplitude level for the emission of \formula{$\kA$} from a dipole-source spanned by \formula{$\pp$} and \formula{$\kB$}, but this time including the effect due to the fermion spin as well.
Not only this assures gauge invariance, but also gives a scale for the appropriate logarithm if the integration would be completed.

At this point, it seems straightforward to interpret expressions like \Expression{dipol} present in \Expression{ImixAB} or \Expression{ImixBA}
as factors for the emission from incoming fermions and with angle distributions controlled by the direction of the other gluon, completing at the same time the second arm of the dipole.
%
This interpretation is based mainly on the visual appearance of the expressions, ignoring the structure of singular terms and/or interferences, and therefore might be misleading.
On the other hand, we want to note that our observations are similar to the ones which can be found in the literature, where indeed such interpretation was performed for the sake of constructing parton shower models, even though usually some approximations were used then.
%

Let us look at \Expression{dipol}, as present \eg\ in the first line of \Expression{ImixAB}.
Independently whether we interpret this factor as the description of the emission of a gluon or not, the momentum entering \formula{$\JJ$} from the left side is \formula{$\pp-\kA$}.
If the collinear limit can be used, and thus to a good approximation \formula{$\JJ$} varies slowly with an eventual virtuality \formula{$(\pp-\kA)^2$}, then the effective intermediate state of a fermion of momentum \formula{$\pp-\kA$} can be used to simplify the description.
A similar line of arguments is used in \cite{Was:1994kg} (and since long
in any formulation of the PDF evolution) and we will not repeat it here.
Note that the intermediate effective gluon definition introduced in \Section{MixedR} is valid all over phase space, even in phase space regions where it does not have any physical meaning anymore.
We do not have such descriptions for an intermediate effective fermion available at this moment: we would have to stick to the low virtuality limit, and thus have to introduce an approximation.

Let us shift our attention to the second line of \Expression{ImixAB}.
For the simplified picture of the previous paragraph to work, one of the effective incoming momenta must be a collinear projection of \formula{$\pp-\kA-\kB$} (or \formula{$\qq-\kA-\kB$}), while the other one simply remains \formula{$\qq$} or \formula{$\pp$}.
No ambiguity appears in this respect.
Regarding the factors \formula{$(\inp{\qq}{\kA})/(\inp{\qq}{\kA}+\inp{\qq}{\kB}-\inp{\kB}{\kA})$} and \formula{$(\inp{\pp}{\kA})/(\inp{\pp}{\kA}+\inp{\pp}{\kB}-\inp{\kB}{\kA})$} we want to mention that, in case of collinear configurations, they are indispensable for the redefinition of the spinor normalization used in the definition of evolution kernels.
As we do not want to limit ourselves to such configurations, but want to use expressions valid all over the phase space, definitions of spinors for intermediate, seemingly on mass-shell, fermion states should remain valid 
everywhere as well.
We do not  have such definition available.
%
%
We want to conclude the discussion about the interpretation of the emission-like factors with the remark that outside of some specific regions of phase space, the interferences between contributions of different chains (as given by \eg\ \Expression{bastart}) can be of the same order as their squares, and thus cannot be ignored.

Let us now turn our attention to \Expression{Imixcom}.
It is rather tempting to interpret its $\pp$- and $\qq$-dependent parts as the real-emission contributions to the running of the coupling constant (of the single gluon emission).
The most singular parts of these contributions cancel each other partly
as we can see in \Equation{Eq1448}.
Such a partial result may already provide a hint on the possible optimal choice of scale to be used as an argument of the running coupling constant.

The role of the remaining parts of the amplitude, expression \Expression{Imixant}, is less clear, although it resembles the running coupling constant part.
It may equally well be interpreted as a genuine second-order part of the amplitude, which can not be interpreted in the language suitable for resummation at all.
Note that, from the point of view of $pq\to\kA\kB\JJ$ kinematics, this contribution is less singular than any of the previously discussed ones.

In the next section, we verify how our somewhat naive interpretation will fare in regions of phase space where iterative descriptions such as BFKL \cite{Kuraev:1977fs,Balitsky:1978ic}, DGLAP \cite{Altarelli:1977zs,Gribov:1972ri,Dokshitzer:1977sg} or CCFM \cite{Catani:1989sg,Catani:1989yc,Ciafaloni:1987ur} are valid and are known to be useful.
%

\section{Picture of ordered gluon emission in dipole language\label{SecCases}}
In the previous chapters we have collected several forms for exact spin amplitudes.
%
%
In the following subsections, we concentrate on cases of special limits, usually associated with some type of ordering.
We will attempt to define amplitudes which are dominant in certain regions of phase space, but nonetheless valid {\em all\/} over phase space, and which differ from the complete amplitude only by gauge invariant and analytically available expressions.
In every presentation of the following subsections, we will start from the exact spin amplitudes as given in Section \ref{TAB} and \ref{MixedR}, but of course our aim is to define correspondence between our exact fixed order result and approximate ones, which are at the heart of successful Monte Carlo programs and are valid at higher orders as well.
%
%

\subsection{\formula{$x$} ordering and soft gluon limit (BFKL)}
Let us restrict our attention to the region of phase space where \formula{$\sqrt{s} \gg \kA^0 \gg \kB^0$} and \formula{$\kA^0 \kB^0 \simeq$} \formula{$\inp{\kA}{\kB}$} (all in the reaction rest frame).
As a consequence, we can use the conditions
 \formula{$\inp{\pp}{\kA}\gg\inp{\pp}{\kB}\gg\inp{\kA}{\kB}$} and/or \formula{$\inp{\qq}{\kA}\gg\inp{\qq}{\kB}\gg\inp{\kA}{\kB}$} for the localization of dominant terms.
Such a choice is consistent with the BFKL approximation \cite{Kuraev:1977fs,Balitsky:1978ic}.
Under these constraints, the \formula{$\Ta\Tb$}-part \Expression{ImixAB} of our expression reduces to
%
\begin{equation}
\ImixAB
=\Jsl
\bigg(
 \frac{\inp{\pp}{\eA}}{\inp{\pp}{\kA}}
-\frac{\inp{\kB}{\eA}}{\inp{\kB}{\kA}}
\bigg)
\bigg(
\frac{\inp{\kA}{\eB}}{\inp{\kA}{\kB}}
- \frac{\inp{\qq}{\eB}}{\inp{\qq}{\kB}}
\bigg)
+
\Jsl
\bigg(
  \frac{\inp{\qq}{\eA}}{\inp{\qq}{\kA}}
 -\frac{\inp{\kB}{\eA}}{\inp{\kB}{\kA}}
\bigg)
\bigg(
  \frac{\inp{\qq}{\eB}}{\inp{\qq}{\kB}}
 -\frac{\inp{\kA}{\eB}}{\inp{\kA}{\kB}}
\bigg)
~.
\label{Eq2037}
\end{equation}
%
All other parts  are  smaller.
For the \formula{$\Tb\Ta$}-part \Expression{ImixBA} we get
%
\begin{equation}
\ImixBA
=
\Jsl
\bigg(
 \frac{\inp{\pp}{\eB}}{\inp{\pp}{\kB}}
-\frac{\inp{\kA}{\eB}}{\inp{\kA}{\kB}}
\bigg)
\bigg(
\frac{\inp{\kB}{\eA}}{\inp{\kB}{\kA}}
-\frac{\inp{\qq}{\eA}}{\inp{\qq}{\kA}}
\bigg)
+
\Jsl
\bigg(
  \frac{\inp{\pp}{\eB}}{\inp{\pp}{\kB}}
 -\frac{\inp{\kA}{\eB}}{\inp{\kA}{\kB}}
\bigg)
\bigg(
  \frac{\inp{\pp}{\eA}}{\inp{\pp}{\kA}}
 -\frac{\inp{\kB}{\eA}}{\inp{\kB}{\kA}}
\bigg)
~.
\label{Eq2070}
\end{equation}
%
After some short manipulations, and since the contributions from \Equation{Imixcom} and \Equation{Imixant} are also negligible thanks to the conditions, the full amplitude reduces to:
%
\begin{multline}
\Mab_{\mathit{BFKL}}
=
\frac{1}{2}
\,\bar{v}(\pp)\Jsl\,u(\qq)
\bigg[
\Ta\Tb
\bigg(
 \frac{\inp{\qq}{\eA}}{\inp{\qq}{\kA}}
-\frac{\inp{\pp}{\eA}}{\inp{\pp}{\kA}}
\bigg)
\bigg(
 \frac{\inp{\qq}{\eB}}{\inp{\qq}{\kB}}
-\frac{\inp{\kA}{\eB}}{\inp{\kA}{\kB}}
\bigg)
\\
+
\Tb\Ta
\bigg(
  \frac{\inp{\pp}{\eA}}{\inp{\pp}{\kA}}
- \frac{\inp{\qq}{\eA}}{\inp{\qq}{\kA}}
\bigg)
\bigg(
  \frac{\inp{\pp}{\eB}}{\inp{\pp}{\kB}}
 -\frac{\inp{\kA}{\eB}}{\inp{\kA}{\kB}}
\bigg)
\bigg]
~.
\label{MBFKL}
\end{multline}
%
A picture of linked dipoles is manifest in this expression.
%
It is also worth mentioning that \Expression{Imixcom}, suspected to contribute to the running of the coupling constant, can be neglected under our conditions.
%
%
The obtained amplitude \Expression{MBFKL}, is a well defined part of the exact one,
and at the same time is consistent with the BFKL approximation.
Note the absence  of
%
\begin{equation}
\Jsl\,\frac{\inp{\kA}{\eB}}{\inp{\kA}{\kB}}\,\frac{\inp{\kB}{\eA}}{\inp{\kA}{\kB}}
\label{virtgluP}
\end{equation}
%
in \Expression{MBFKL}.
It is a clear manifestation of large destructive interferences between gauge invariant parts of our expressions for the amplitude given in Sections \ref{TAB} and \ref{MixedR}.
A crude distribution, of some hypothetical Monte Carlo, based on the incoherent sum of \Equation{Eq2037} and \Equation{Eq2070} would possibly be less efficient than one based on \Expression{MBFKL}.
%
At the level of double-emission taken alone, this observation  is of not much interest of course.

Finally, let us mention that if the large \formula{$\ncol$} limit is used, the interference between two distinct color contributions to \Expression{MBFKL} in the amplitude squared and averaged over spin degrees of freedom is suppressed by a factor \formula{$\ncol^2$}.
It is a direct consequence of the fact that
%
\begin{equation}
\sum_{a,b} \Tr(\Ta\Tb \Ta\Tb)
=
\frac{-\ncol}{4}\bigg(1-\frac{1}{\ncol^2}\bigg)
\end{equation}
%
and
%
\begin{equation}
\sum_{a,b} \Tr(\Ta\Tb \Tb\Ta)
=
\frac{\ncol^3}{4}\bigg(1-\frac{1}{\ncol^2}\bigg)^2
~.
\end{equation}
%

That completes the study of this kinematical configuration.
We have explicitly truncated the complete amplitude to obtain an expression consisting of two incoherent dipole chains, which are easy to implement into a Monte Carlo program incorporated with exact Lorentz invariant phase space as  in QED \cite{Barberio:1994qi,Nanava:2006vv} for the solution based on the exact matrix element and full
phase space coverage.

Note that the truncated amplitude is given in the same kinematical formulation as the complete exact amplitude for two-gluon emission.
It can be calculated at any point in phase space, and not only where the approximation is justified.
That is why the re-installation of the exact amplitude into a calculation based on sufficiently refined dipole language can be accommodated for by a well-defined
weight, namely the ratio of matrix elements squared.

We can conclude that this part of the exercise is finished as far as the analysis of two-gluon emission amplitudes is concerned.
It can serve as a basis for studies of algorithms where the soft-gluon
approximation is used as first step in event construction.
%
%
%
%
%
The next step of our work must include other processes of double-parton
emissions taken at tree level.
Only then we can be sure that the method as used in the PHOTOS Monte Carlo can be of any interest for QCD.
We want to note, that the picture of linked dipoles is used with great success in Ariadne parton shower \cite{Lonnblad:1992tz} since quite some time now.
%

\subsubsection{ Two soft gluons, general case}
Let us modify the previously discussed case now.
We will demand
that \formula{$\sqrt{s} \gg \kA^0  ({\rm or} \;\kB^0)$} but we will   not ignore
regions of phase space where
\formula{$\kA^0 \kB^0 \gg \inp{\kA}{\kB}$} (this condition is to be
taken again in the reaction frame).
Under such conditions we may neglect all terms like \formula{$(\esA\ksA)/(2\inp{\pp}{\kA})$}.
If the scalar product \formula{$\inp{\kA}{\kB}$} appears in a propagator denominator together with larger terms, it can be dropped from that denominator.
Thanks to the above approximation, we find that \Expression{ImixAB}, \Expression{ImixBA}, \Expression{Imixcom}  and \Expression{Imixant} simplify again
%
\begin{align}
\ImixAB
&=
\Jsl
\bigg(
 \frac{\inp{\pp}{\eA}}{\inp{\pp}{\kA}}
-\frac{\inp{\kB}{\eA}}{\inp{\kB}{\kA}}
\bigg)
\bigg(
 \frac{\inp{\kA}{\eB}}{\inp{\kA}{\kB}}
-\frac{\inp{\qq}{\eB}}{\inp{\qq}{\kB}}
\bigg)
\nonumber\\
&+
\Jsl 
\,\frac{\inp{\pp}{\kB}}{\inp{\pp}{\kA}+\inp{\pp}{\kB}}
\bigg(
  \frac{\inp{\pp}{\eA}}{\inp{\pp}{\kA}}
 -\frac{\inp{\kB}{\eA}}{\inp{\kB}{\kA}}
\bigg)
\bigg(
  \frac{\inp{\pp}{\eB}}{\inp{\pp}{\kB}}
 -\frac{\inp{\kA}{\eB}}{\inp{\kA}{\kB}}
\bigg)
\\
&+
\Jsl 
\,\frac{\inp{\qq}{\kA}}{\inp{\qq}{\kA}+\inp{\qq}{\kB}}
\bigg(
  \frac{\inp{\qq}{\eA}}{\inp{\qq}{\kA}}
 -\frac{\inp{\kB}{\eA}}{\inp{\kB}{\kA}}
\bigg)
  \bigg(\frac{\inp{\qq}{\eB}}{\inp{\qq}{\kB}}
 -\frac{\inp{\kA}{\eB}}{\inp{\kA}{\kB}}
\bigg)
\nonumber
\end{align}
%
\begin{align}
\ImixBA
&=
\Jsl
\bigg(
 \frac{\inp{\pp}{\eB}}{\inp{\pp}{\kB}}
-\frac{\inp{\kA}{\eB}}{\inp{\kA}{\kB}}
\bigg)
\bigg(
 \frac{\inp{\kB}{\eA}}{\inp{\kB}{\kA}}
-\frac{\inp{\qq}{\eA}}{\inp{\qq}{\kA}}
\bigg)
\nonumber\\
&+
\Jsl 
\,\frac{\inp{\pp}{\kA}}{\inp{\pp}{\kB}+\inp{\pp}{\kA}}
\bigg(
  \frac{\inp{\pp}{\eB}}{\inp{\pp}{\kB}}
 -\frac{\inp{\kA}{\eB}}{\inp{\kA}{\kB}}
\bigg)
\bigg(
  \frac{\inp{\pp}{\eA}}{\inp{\pp}{\kA}}
 -\frac{\inp{\kB}{\eA}}{\inp{\kB}{\kA}}
\bigg)
\\
&+
\Jsl 
\,\frac{\inp{\qq}{\kB}}{\inp{\qq}{\kB}+\inp{\qq}{\kA}}
\bigg(
  \frac{\inp{\qq}{\eB}}{\inp{\qq}{\kB}}
 -\frac{\inp{\kA}{\eB}}{\inp{\kA}{\kB}}
\bigg)
\bigg(
  \frac{\inp{\qq}{\eA}}{\inp{\qq}{\kA}}
  -\frac{\inp{\kB}{\eA}}{\inp{\kB}{\kA}}
\bigg)
\nonumber
\end{align}
%
\begin{equation}
\Imixcom
=
\frac{1}{2}
\,\Jsl 
\bigg(
  \frac{\inp{\pp}{\kA}-\inp{\pp}{\kB}}{\inp{\pp}{\kB}+\inp{\pp}{\kA}}
 +\frac{\inp{\qq}{\kB}-\inp{\qq}{\kA}}{\inp{\qq}{\kB}+\inp{\qq}{\kA}}
\bigg)
\bigg(
  \frac{\inp{\kB}{\eA}}{\inp{\kB}{\kA}}\,\frac{\inp{\kA}{\eB}}{\inp{\kB}{\kA}}
      -\frac{\inp{\eB}{\eA}}{\inp{\kB}{\kA}}
\bigg).
\label{barunx}
\end{equation}
%
Both \formula{$\ImixAB$} and \formula{$\ImixBA$} are now build from scalar factors and \formula{$\Jsl$}.
The \formula{$\Imixcom$}-part, potentially  a contribution to running coupling constant, remains this time.
Recall that it is free of singularities when the gluons become 
collinear thanks to a cancellation between contributions of the quark and anti-quark lines.
All other parts, and in particular all parts of $\Imixant$, could be dropped out as non-dominant.
Let us study the terms surviving the approximation.
After minor manipulations, we find that the approximated amplitude can be written as the sum of two terms, the first of which is given by
%
\begin{multline}
\Mab_{\mathit{BFKL'}}
=
\frac{1}{2}\,\bar{v}(\pp)\Jsl\,u(\qq)
\\
\times
\bigg\{
\phantom{x+}
\Ta\Tb
\bigg[
\frac{\inp{\pp}{\kB}}{\inp{\pp}{\kA}+\inp{\pp}{\kB}}
\,\bigg(\frac{\inp{\pp}{\eA}}{\inp{\pp}{\kA}}-\frac{\inp{\kB}{\eA}}{\inp{\kB}{\kA}}\bigg)
  \bigg(\frac{\inp{\pp}{\eB}}{\inp{\pp}{\kB}}-\frac{\inp{\qq}{\eB}}{\inp{\qq}{\kB}}\bigg)
\phantom{xxxxxxxx}
\\
\phantom{xxxxxxxxxxxxxxxxx}
+\frac{\inp{\qq}{\kA}}{\inp{\qq}{\kA}+\inp{\qq}{\kB}}\,
\bigg(\frac{\inp{\qq}{\eA}}{\inp{\qq}{\kA}}-\frac{\inp{\pp}{\eA}}{\inp{\pp}{\kA}}\bigg)
\bigg(\frac{\inp{\qq}{\eB}}{\inp{\qq}{\kB}}-\frac{\inp{\kA}{\eB}}{\inp{\kA}{\kB}}\bigg)
\bigg]
\\
+\Tb\Ta
\bigg[
\frac{\inp{\pp}{\kA}}{\inp{\pp}{\kB}+\inp{\pp}{\kA}}
\,\bigg(\frac{\inp{\pp}{\eB}}{\inp{\pp}{\kB}}-\frac{\inp{\kA}{\eB}}{\inp{\kA}{\kB}}\bigg)
  \bigg(\frac{\inp{\pp}{\eA}}{\inp{\pp}{\kA}}-\frac{\inp{\qq}{\eA}}{\inp{\qq}{\kA}}\bigg)
\phantom{xxxx}
\\
\phantom{xxxxxxxxxxxxxxxxx}
+\frac{\inp{\qq}{\kB}}{\inp{\qq}{\kB}+\inp{\qq}{\kA}}
\,\bigg(\frac{\inp{\qq}{\eB}}{\inp{\qq}{\kB}}-\frac{\inp{\pp}{\eB}}{\inp{\pp}{\kB}}\bigg)
  \bigg(\frac{\inp{\qq}{\eA}}{\inp{\qq}{\kA}}-\frac{\inp{\kB}{\eA}}{\inp{\kB}{\kA}}\bigg)
\bigg]
\\
+\frac{[\Ta,\Tb]}{4}
\bigg(
  \frac{\inp{\pp}{\kA}-\inp{\pp}{\kB}}{\inp{\pp}{\kA}+\inp{\pp}{\kB}}
+ \frac{\inp{\qq}{\kB}-\inp{\qq}{\kA}}{\inp{\qq}{\kA}+\inp{\qq}{\kB}}
\bigg)
\phantom{xxxxxxxxxxxxxx}
\\
\phantom{xxxxxx}
\times
\bigg[
 \frac{\inp{\kA}{\eB}}{\inp{\kA}{\kB}}
 \bigg(\frac{\inp{\pp}{\eA}}{\inp{\pp}{\kA}}+\frac{\inp{\qq}{\eA}}{\inp{\qq}{\kA}}\bigg)
+\bigg(\frac{\inp{\pp}{\eB}}{\inp{\pp}{\kB}}+\frac{\inp{\qq}{\eB}}{\inp{\qq}{\kB}}\bigg)
 \frac{\inp{\kB}{\eA}}{\inp{\kB}{\kA}}
\\
\phantom{xxxxxxxxxxxxxxxxxxxxxxxxxxxx}
-\frac{\inp{\pp}{\eB}}{\inp{\pp}{\kB}}\,\frac{\inp{\qq}{\eA}}{\inp{\qq}{\kA}}
-\frac{\inp{\pp}{\eA}}{\inp{\pp}{\kA}}\,\frac{\inp{\qq}{\eB}}{\inp{\qq}{\kB}}
-2\,\frac{\inp{\eB}{\eA}}{\inp{\kB}{\kA}}
\bigg]
\bigg\}
~.
\label{MBFKLp}
\end{multline}
%
The expression differs from the one discussed in the previous subsection.
More terms, of the linked dipole type, survive.
They are weighted by spin-and color-independent factors.
Some remnants from \formula{$\ImixAB$} and \formula{$\ImixBA$} which survive the approximation are combined with the surviving pieces from \formula{$\Imixcom$}.
Together, these parts form the contribution to the coefficient of \formula{$[\Ta,\Tb]$}, for the contribution to the running coupling constant.
However, such a manipulation may considered to be somewhat supeficial.
Note that it concerns terms proportional to 
%
\begin{equation}
\bigg(
  \frac{\inp{\pp}{\kA}-\inp{\pp}{\kB}}{\inp{\pp}{\kA}+\inp{\pp}{\kB}}
+ \frac{\inp{\qq}{\kB}-\inp{\qq}{\kA}}{\inp{\qq}{\kA}+\inp{\qq}{\kB}}
\bigg)
~,
\end{equation}
%
which approaches zero when $\kA$ and $\kB$ become collinear.
It is of course essential in this that we do not separate the expression into a 
$\pp$-and a $\qq$-part.
The second term is given by
%
\begin{multline}
\Mab_{\mathit{BFKL'remnant}}
=
\frac{1}{8}
\,\bar{v}(\pp)\Jsl\,u(\qq) \bigg(
  \frac{\inp{\pp}{\kA}-\inp{\pp}{\kB}}{\inp{\pp}{\kA}+\inp{\pp}{\kB}}
+ \frac{\inp{\qq}{\kB}-\inp{\qq}{\kA}}{\inp{\qq}{\kA}+\inp{\qq}{\kB}}
\bigg)
\\
\times
\bigg\{
\phantom{+}
\Ta\Tb
\bigg[
 \bigg( \frac{\inp{\pp}{\eA}}{\inp{\pp}{\kA}} - \frac{\inp{\qq}{\eA}}{\inp{\qq}{\kA}} \bigg)
 \bigg( \frac{\inp{\kA}{\eB}}{\inp{\kB}{\kA}} - \frac{\inp{\pp}{\eB}}{\inp{\qq}{\kB}} \bigg)
+\bigg( \frac{\inp{\qq}{\eA}}{\inp{\qq}{\kA}} - \frac{\inp{\kB}{\eA}}{\inp{\kA}{\kB}} \bigg)
 \bigg( \frac{\inp{\pp}{\eB}}{\inp{\pp}{\kB}} - \frac{\inp{\qq}{\eB}}{\inp{\qq}{\kB}} \bigg)
\bigg]
\\
\phantom{xxx}
-\Tb\Ta
\bigg[
 \bigg( \frac{\inp{\pp}{\eB}}{\inp{\pp}{\kB}} - \frac{\inp{\qq}{\eB}}{\inp{\qq}{\kB}} \bigg)
 \bigg( \frac{\inp{\kB}{\eA}}{\inp{\kA}{\kB}} - \frac{\inp{\pp}{\eA}}{\inp{\qq}{\kA}} \bigg)
+\bigg( \frac{\inp{\qq}{\eB}}{\inp{\qq}{\kB}} - \frac{\inp{\kA}{\eB}}{\inp{\kB}{\kA}} \bigg)
 \bigg( \frac{\inp{\pp}{\eA}}{\inp{\pp}{\kA}} - \frac{\inp{\qq}{\eA}}{\inp{\qq}{\kA}} \bigg)
\bigg]
\\
+\,[\Ta,\Tb]
 \bigg( \frac{\inp{\pp}{\eA}}{\inp{\pp}{\kA}} - \frac{\inp{\qq}{\eA}}{\inp{\qq}{\kA}} \bigg)
 \bigg( \frac{\inp{\pp}{\eB}}{\inp{\pp}{\kB}} - \frac{\inp{\qq}{\eB}}{\inp{\qq}{\kB}} \bigg)
\phantom{ixxxxxxxxxxxxxxxxxxxxxx}
\bigg\}
~.
\label{MBFKLremn}
\end{multline}
%
It is again non-leading
but can be interpreted as a correction to the dipole
picture.

Finally let us observe that it is straightforward to reproduce the amplitude of the previous subsection.
If we would assume that either \formula{$\kA \ll \kB$} or \formula{$\kB \ll \kA$},
then all spin and color independent factors reduce  either to zero or to one.
%
%

\subsection{ \formula{$\pp_T$} ordering or DGLAP}
In this subsection, we use the type of phase space organization (or in fact approximation) that is called $\pp_T$ ordering.
As in the previous case, it can be formulated in terms of properties of Lorentz invariant expressions, allowing for an easy identification of the parts of the amplitudes which may be simply dropped out.
In the following, we will discuss the two kinematical cases of emissions into one, or two (opposite) hemispheres.
We will devote a separate subsection to the discussion of the running coupling constant contribution.
%

\subsubsection{Dominant parts for emissions in one hemisphere} \label{chap:-p-t-j}
We assume first that \formula{$\inp{\pp}{\kA}\gg\inp{\pp}{\kB}$} or \formula{$\inp{\qq}{\kA}\gg\inp{\qq}{\kB}$}, but we accept configurations where \formula{$\inp{\pp}{\kA}\simeq\inp{\kA}{\kB}$} or \formula{$\inp{\qq}{\kA}\simeq\inp{\kA}{\kB}$}.
In practice, such conditions means that \formula{$\kB$}, \formula{$\pp$} and \formula{$\qq$} are basically parallel to each other from the point of view 
 of the direction of \formula{$\kA$}.
The `macroscopic' size of \formula{$\inp{\kA}{\kB}$} implicitly removes the phase space regions where the gluons become collinear.
Such a case must be treated separately.
Under our conditions, the \formula{$\Ta\Tb$} part of the amplitude takes the form:
%
\begin{align}
\ImixAB
&=
\bigg(
 \frac{\inp{\pp}{\eA}}{\inp{\pp}{\kA}}
-\frac{\inp{\kB}{\eA}}{\inp{\kB}{\kA}}
-\frac{\esA\ksA}{2\inp{\pp}{\kA}}
\bigg)
\Jsl
\bigg(
 \frac{\ksB\esB}{2\inp{\qq}{\kB}}
+\frac{\inp{\kA}{\eB}}{\inp{\kA}{\kB}}
-\frac{\inp{\qq}{\eB}}{\inp{\qq}{\kB}}
\bigg)
\nonumber\\
&+
\Jsl 
\,\frac{\inp{\qq}{\kA}}{\inp{\qq}{\kA}-\inp{\kA}{\kB}}
\,\bigg(\frac{\inp{\qq}{\eA}}{\inp{\qq}{\kA}}-\frac{\inp{\kB}{\eA}}{\inp{\kB}{\kA}}
        -\frac{\ksA\esA}{2\inp{\qq}{\kA}}\bigg)
  \bigg(\frac{\inp{\qq}{\eB}}{\inp{\qq}{\kB}}-\frac{\inp{\kA}{\eB}}{\inp{\kA}{\kB}}
        -\frac{\ksB\esB}{2\inp{\qq}{\kB}}\bigg)
~.
\end{align}
%
Analogously,  the \formula{$\Tb\Ta$} part reads:
%
\begin{align}
\ImixBA
&=
\bigg(
 \frac{\inp{\pp}{\eB}}{\inp{\pp}{\kB}}
-\frac{\inp{\kA}{\eB}}{\inp{\kA}{\kB}}
-\frac{\esB\ksB}{2\inp{\pp}{\kB}}
\bigg)
\Jsl
\bigg(
 \frac{\ksA\esA}{2\inp{\qq}{\kA}}
+\frac{\inp{\kB}{\eA}}{\inp{\kB}{\kA}}
-\frac{\inp{\qq}{\eA}}{\inp{\qq}{\kA}}
\bigg)
\nonumber\\
&+
\frac{\inp{\pp}{\kA}}{\inp{\pp}{\kA}-\inp{\kB}{\kA}}
\,\bigg(\frac{\inp{\pp}{\eB}}{\inp{\pp}{\kB}}-\frac{\inp{\kA}{\eB}}{\inp{\kA}{\kB}}
        -\frac{\esB\ksB}{2\inp{\pp}{\kB}}\bigg)
  \bigg(\frac{\inp{\pp}{\eA}}{\inp{\pp}{\kA}}-\frac{\inp{\kB}{\eA}}{\inp{\kB}{\kA}}
        -\frac{\esA\ksA}{2\inp{\pp}{\kA}}\bigg)
\Jsl
~.
\end{align}
%
It is only a matter of a short manipulations for the \formula{$\Ta\Tb$} amplitude part to take the form: 
%
\begin{multline}
\ImixAB
=  - \frac{\inp{\kA}{\kB}}{\inp{\qq}{\kA}-\inp{\kA}{\kB}}
\bigg(
 \frac{\inp{\pp}{\eA}}{\inp{\pp}{\kA}}
-\frac{\inp{\kB}{\eA}}{\inp{\kB}{\kA}}
-\frac{\esA\ksA}{2\inp{\pp}{\kA}}
\bigg)
\Jsl
\bigg(
 \frac{\ksB\esB}{2\inp{\qq}{\kB}}
+\frac{\inp{\kA}{\eB}}{\inp{\kA}{\kB}}
-\frac{\inp{\qq}{\eB}}{\inp{\qq}{\kB}}
\bigg)
\\
+
\,\frac{\inp{\qq}{\kA}}{\inp{\qq}{\kA}-\inp{\kA}{\kB}}
\biggl\{\Jsl
\,\bigg(\frac{\inp{\qq}{\eA}}{\inp{\qq}{\kA}}-\frac{\inp{\kB}{\eA}}{\inp{\kB}{\kA}}
        -\frac{\ksA\esA}{2\inp{\qq}{\kA}}\bigg)
-\bigg(
 \frac{\inp{\pp}{\eA}}{\inp{\pp}{\kA}}
-\frac{\inp{\kB}{\eA}}{\inp{\kB}{\kA}}
-\frac{\esA\ksA}{2\inp{\pp}{\kA}}
\bigg)
\Jsl
\biggr\}
\phantom{xxx}
\\
\times
  \bigg(\frac{\inp{\qq}{\eB}}{\inp{\qq}{\kB}}-\frac{\inp{\kA}{\eB}}{\inp{\kA}{\kB}}
        -\frac{\ksB\esB}{2\inp{\qq}{\kB}}\bigg)
~,
\label{DG2x}
\end{multline}
%
and analogously, for the \formula{$\Tb\Ta$} part we find:
%
\begin{multline}
\ImixBA
= -\frac{\inp{\kA}{\kB}}{\inp{\pp}{\kA}-\inp{\kB}{\kA}}
\bigg(
 \frac{\inp{\pp}{\eB}}{\inp{\pp}{\kB}}
-\frac{\inp{\kA}{\eB}}{\inp{\kA}{\kB}}
-\frac{\esB\ksB}{2\inp{\pp}{\kB}}
\bigg)
\Jsl
\bigg(
 \frac{\ksA\esA}{2\inp{\qq}{\kA}}
+\frac{\inp{\kB}{\eA}}{\inp{\kB}{\kA}}
-\frac{\inp{\qq}{\eA}}{\inp{\qq}{\kA}}
\bigg)
\\
+
\frac{\inp{\pp}{\kA}}{\inp{\pp}{\kA}-\inp{\kB}{\kA}}
\,\bigg(\frac{\inp{\pp}{\eB}}{\inp{\pp}{\kB}}-\frac{\inp{\kA}{\eB}}{\inp{\kA}{\kB}}
        -\frac{\esB\ksB}{2\inp{\pp}{\kB}}\bigg)
\phantom{xxxxxxxxxxxxxxxxxxxxxxxxxxxxx}
\\
\times
\biggl\{
  \bigg(\frac{\inp{\pp}{\eA}}{\inp{\pp}{\kA}}-\frac{\inp{\kB}{\eA}}{\inp{\kB}{\kA}}
        -\frac{\esA\ksA}{2\inp{\pp}{\kA}}\bigg)
\Jsl +
\Jsl
\bigg(
 \frac{\ksA\esA}{2\inp{\qq}{\kA}}
+\frac{\inp{\kB}{\eA}}{\inp{\kB}{\kA}}
-\frac{\inp{\qq}{\eA}}{\inp{\qq}{\kA}}\bigg)
\biggr\}
~.
\label{DG4x}
\end{multline}

The above expressions can easily be understood in the language of evolution kernels.
The factors \formula{$(\inp{\qq}{\kA})/(\inp{\qq}{\kA}-\inp{\kB}{\kA})$} and \formula{$(\inp{\pp}{\kA})/(\inp{\pp}{\kA}-\inp{\kB}{\kA})$} can be understood as redefinition of the spinor normalization, again similarly as explained in \cite{Was:1994kg}.
Note that the contents of the curly brackets in the last two formulas are free of terms proportional to 
\formula{$(\inp{\kA}{\eB})/(\inp{\kB}{\kA})$}  and \formula{$(\inp{\kB}{\eA})/(\inp{\kB}{\kA})$}.
They represent simple dipoles spanned on the \formula{$\pp$},\formula{$\qq$}-pair.
In contrast with the previous section, correcting terms with \formula{$\ksA$} and/or \formula{$\ksB$} in numerator remain.
This is closely related to the necessity/possibility to introduce effective momenta flowing into \formula{$\JJ$}, once the language of PDFs is introduced.
Note, that thanks to the \formula{$\pp_T$} ordering there are only two chains of dipoles left, similarly as it was in the BFKL-case.

This time the coefficient \Expression{Imixcom1} for the running coupling constant seems to survive in its near complete form.
In our limit and because of cancellations it develops an extra power of \formula{$\inp{\kA}{\kB}$} in the eikonal part, and a factor of \formula{$\sqrt{\inp{\kA}{\kB}}$} in the terms proportional to \formula{$\kAB$}.
This will cancel the singularity of the virtual gluon once the amplitude is squared and a partial integration over the phase space is performed.
Because of this, an interpretation in the language of the running coupling constant is prevented.
It can be treated as a non-singular correction, and thus ignored.
At this point, the freedom of choice for the arguments of the running coupling constant remains unrestricted.
%
%
Note also, that remnants from the dipole parts did not migrate into the coupling constant expression, in contrast with the situation for BFKL in the previous section.
This observation can be  justified further.
If we return to the soft gluon case of the previous section, and apply the \formula{$\pp_T$} ordering as well, then the running coupling constant contribution of \Equation{MBFKLp} will become suppressed.
We will return to this point later in \Section{Sec533}.

For now, we can conclude that our amplitude, with  properties consistent with \formula{$\pp_T$} ordering, is  more compact than the exact one. 
It is gauge invariant and valid all over phase space.
Such a simple, amplitude-level, expression can be used at the intermediate step for the definition of a parton shower algorithm, or to better understand the already existing ones.
As in the previous BFKL case, an explicit form for the weight necessary to reinstall the exact distribution based on the two-gluon amplitude is available.
It was mentioned already earlier how to identify emission kernels already at the spin amplitude level.
It seems that the explanations included in \cite{Was:1994kg} can be applied here as well.
For a further continuation of our discussion, the language of the truncated 
exact spin amplitudes is probably too restrictive: a particular choice for the direction of \formula{$\kA$} and/or \formula{$\kB$} in the vicinity of \formula{$\pp$} (or \formula{$\qq$}) must be made, and generates a multitude of options. 
Furthermore, a definition of intermediate states for effective fermions valid all over the phase space would be necessary.
%
%

\subsubsection{Dominant parts for simultaneous emissions from $\pp$ and $\qq$}

%
Now, let us  assume that \formula{$\inp{\pp}{\kA}\gg\inp{\pp}{\kB}$} and \formula{$\inp{\qq}{\kB}\gg\inp{\qq}{\kA}$}.
Under such conditions, \formula{$\kB$} is basically parallel to \formula{$\pp$}, and \formula{$\kA$} to \formula{$\qq$}.
It is obvious that contribution \Expression{bastart} dominates over all other contributions of the complete spin amplitude.
If in addition \formula{$\inp{\pp}{\kA}\gg\inp{\qq}{\kB}$}, then we can replace \formula{$(\inp{\kB}{\eA})/(\inp{\kB}{\kA})$} with \formula{$(\inp{\pp}{\eA})/(\inp{\pp}{\kA})$} and get
%
\begin{equation}
\IBA
=
\bigg(
 \frac{\inp{\pp}{\eB}}{\inp{\pp}{\kB}}
-\frac{\inp{\kA}{\eB}}{\inp{\kA}{\kB}}
-\frac{\esB\ksB}{2\inp{\pp}{\kB}}
\bigg)
\Jsl
\bigg(
 \frac{\ksA\esA}{2\inp{\qq}{\kA}}
+\frac{\inp{\pp}{\eA}}{\inp{\pp}{\kA}}
-\frac{\inp{\qq}{\eA}}{\inp{\qq}{\kA}}
\bigg)
~.
\end{equation}
%
This expression is again of the dipole form.
At the same time it is part of the expression discussed previously.
This means, that the replacement of \formula{$(\inp{\kB}{\eA})/(\inp{\kB}{\kA})$} with \formula{$(\inp{\pp}{\eA})/(\inp{\pp}{\kA})$}, which is potentially dangerous and valid only in this particular region of phase space, is not necessary,
except for our proof here.
Difficulties, as the ones to be discussed in \Section{Sec:ccfm}, can be avoided.

%

\subsubsection{Case when  $\kA$ and $\kB$ may become parallel\label{Sec533}}
So far, we have implicitly excluded the phase space region contributing to the running coupling constant from our discussion.
The \formula{$\pp_T$} ordering makes it unfavorable for the two gluons to become parallel one to another.
Therefore, we have to look at regions of phase space where the virtuality of the gluon may approach zero separately.
For that purpose, we will assume that the overall $\pp_T$ of the virtual gluon is small, but larger than \formula{$\inp{\kA}{\kB}$}.
We have found such an approach in studies presented in \cite{Ermolaev:2007yb,Ermolaev:1981cm}.
We will consider the configuration when \formula{$\inp{\qq}{\kA}\gg\inp{\pp}{\kA}$}, \formula{$\inp{\qq}{\kB}\gg\inp{\pp}{\kB}$},  \formula{$\inp{\pp}{\kA}\gg\inp{\kA}{\kB}$} and \formula{$\inp{\pp}{\kA} \sim \inp{\pp}{\kB}$}.  
Such choice represents the splitting the anti-quark with momentum $\pp$ into a virtual fermion line entering $J$, and a single gluon of small virtuality and (moderately) small \formula{$\pp_T$}.
Under such circumstances, dominant contributions from \Expression{ImixAB}, \Expression{ImixBA}, \Expression{Imixcom} take the form:
%
\begin{align}
\ImixAB
&=
\bigg(
 \frac{\inp{\pp}{\eA}}{\inp{\pp}{\kA}}
-\frac{\inp{\kB}{\eA}}{\inp{\kB}{\kA}}
-\frac{\esA\ksA}{2\inp{\pp}{\kA}}
\bigg)
\Jsl
\bigg(
 \frac{\ksB\esB}{2\inp{\qq}{\kB}}
+\frac{\inp{\kA}{\eB}}{\inp{\kA}{\kB}}
-\frac{\inp{\qq}{\eB}}{\inp{\qq}{\kB}}
\bigg)
\nonumber\\
&+
\frac{\inp{\pp}{\kB}}{\inp{\pp}{\kA}+\inp{\pp}{\kB}}
\,\bigg(\frac{\inp{\pp}{\eA}}{\inp{\pp}{\kA}}-\frac{\inp{\kB}{\eA}}{\inp{\kB}{\kA}}
        -\frac{\esA\ksA}{2\inp{\pp}{\kA}}\bigg)
  \bigg(\frac{\inp{\pp}{\eB}}{\inp{\pp}{\kB}}-\frac{\inp{\kA}{\eB}}{\inp{\kA}{\kB}}
        -\frac{\esB\ksB}{2\inp{\pp}{\kB}}\bigg)
\Jsl
\phantom{xx}
\label{ImixABx}\\
&+
\Jsl 
\,\frac{\inp{\qq}{\kA}}{\inp{\qq}{\kA}+\inp{\qq}{\kB}}
\,\bigg(\frac{\inp{\qq}{\eA}}{\inp{\qq}{\kA}}-\frac{\inp{\kB}{\eA}}{\inp{\kB}{\kA}}
        -\frac{\ksA\esA}{2\inp{\qq}{\kA}}\bigg)
  \bigg(\frac{\inp{\qq}{\eB}}{\inp{\qq}{\kB}}-\frac{\inp{\kA}{\eB}}{\inp{\kA}{\kB}}
        -\frac{\ksB\esB}{2\inp{\qq}{\kB}}\bigg)
\nonumber
\end{align}
%
\begin{align}
\ImixBA
&=
\bigg(
 \frac{\inp{\pp}{\eB}}{\inp{\pp}{\kB}}
-\frac{\inp{\kA}{\eB}}{\inp{\kA}{\kB}}
-\frac{\esB\ksB}{2\inp{\pp}{\kB}}
\bigg)
\Jsl
\bigg(
 \frac{\ksA\esA}{2\inp{\qq}{\kA}}
+\frac{\inp{\kB}{\eA}}{\inp{\kB}{\kA}}
-\frac{\inp{\qq}{\eA}}{\inp{\qq}{\kA}}
\bigg)
\nonumber\\
&+
\frac{\inp{\pp}{\kA}}{\inp{\pp}{\kB}+\inp{\pp}{\kA}}
\,\bigg(\frac{\inp{\pp}{\eB}}{\inp{\pp}{\kB}}-\frac{\inp{\kA}{\eB}}{\inp{\kA}{\kB}}
        -\frac{\esB\ksB}{2\inp{\pp}{\kB}}\bigg)
  \bigg(\frac{\inp{\pp}{\eA}}{\inp{\pp}{\kA}}-\frac{\inp{\kB}{\eA}}{\inp{\kB}{\kA}}
        -\frac{\esA\ksA}{2\inp{\pp}{\kA}}\bigg)
\Jsl
\phantom{xx}
\label{ImixBAx}\\
&+
\Jsl 
\,\frac{\inp{\qq}{\kB}}{\inp{\qq}{\kB}+\inp{\qq}{\kA}}
\,\bigg(\frac{\inp{\qq}{\eB}}{\inp{\qq}{\kB}}-\frac{\inp{\kA}{\eB}}{\inp{\kA}{\kB}}
        -\frac{\ksB\esB}{2\inp{\qq}{\kB}}\bigg)
  \bigg(\frac{\inp{\qq}{\eA}}{\inp{\qq}{\kA}}-\frac{\inp{\kB}{\eA}}{\inp{\kB}{\kA}}
        -\frac{\ksA\esA}{2\inp{\qq}{\kA}}\bigg)
\nonumber
\end{align}
%
\begin{align}
\Imixcom
&=
\frac{1}{2}
\,\Jsl 
\bigg(
  \frac{\inp{\pp}{\kA}-\inp{\pp}{\kB}}{\inp{\pp}{\kA}+\inp{\pp}{\kB}}
+ \frac{\inp{\qq}{\kB}-\inp{\qq}{\kA}}{\inp{\qq}{\kA}+\inp{\qq}{\kB}}
\bigg)
\bigg( \frac{\inp{\kA}{\eB}}{\inp{\kA}{\kB}}\,\frac{\inp{\kB}{\eA}}{\inp{\kA}{\kB}}
      -\frac{\inp{\eA}{\eB}}{\inp{\kA}{\kB}}\bigg)
\nonumber\\
&-
\frac{1}{4}
\,\frac{1}{\inp{\pp}{\kA}+\inp{\pp}{\kB}}
\bigg(
  \frac{ \esA\ksA\esB\ksB -\esB\ksB\esA\ksA }{\inp{\kA}{\kB}}
\bigg)
\Jsl
\label{Imixcomx}\\
&-
\frac{1}{4}
\,\Jsl 
\,\frac{1}{\inp{\qq}{\kA}+\inp{\qq}{\kB}}
\bigg(
  \frac{ \ksA\esA\ksB\esB -\ksB\esB\ksA\esA }{\inp{\kA}{\kB}}
\bigg)
~.
\nonumber
\end{align}
%
Contribution \Expression{Imixant} becomes non-leading as a whole, so we have dropped it out already now.
Obviously our choice of remaining terms is different from the one of \Section{chap:-p-t-j}.
Nonetheless one can observe that \Expression{ImixABx} and \Expression{ImixBAx} are  included in  \Expression{DG2x} and \Expression{DG4x}, if for the latter both possibilities \formula{$\inp{\pp}{\kA}\gg\inp{\pp}{\kB}$} and \formula{$\inp{\pp}{\kA}\ll\inp{\pp}{\kB}$} are added together.
Expression \Expression{Imixcomx} may contribute to the running coupling constant.
%
%
The last line of \Expression{Imixcomx} is non-leading for our choice of kinematical conditions and should be dropped out.
However, it is gauge invariant and rather compact, and it is necessary in case gluons would be collimated with $\qq$.
Also, one should bear in mind that the first line of \Expression{Imixcomx}
is less singular than it may seem, because of a partial cancellation of $\pp$
and $\qq$ dependent terms, which one may be tempted to treat separately.
This is again, as in BFKL case, a consequence of the gauge cancellation, which is not exact as it concerns a virtual gluon.

\subsection{Angle ordering or CCFM style\label{Sec:ccfm}}
The CCFM case is definitely more difficult for our approach than the ones discussed so far.
We have to use angular ordering to select the dominant terms.
Such a choice is less straightforward for the Lorentz invariant representation.
It is thus not possible to simply neglect some terms because they would be explicitly smaller than others.
In fact, it seems that all terms of our expressions for the amplitude will need to be kept.
Some kind of a language exploiting the concept of effective intermediate states is needed, and that is definitely a complication.
Even if approximated amplitudes would be defined, they would differ from the complete ones quite substantially by the presence of these effective intermediate states.
That would definitely be out of scope of the present paper.

On the other hand, we want to remark that the interpretation of the result using some sort of the dipole language will persist in this case as well, as it is already visible in the exact amplitudes.
There will be simply more terms to take into account, that is formulas \Expression{abstart},  \Expression{ab-2},  \Expression{ab-3}, \Expression{bastart}, \Expression{ba-2},  \Expression{ba-3}.
The fermion normalization redefinition factors: \formula{$(\inp{\qq}{\kA})/(\inp{\qq}{\kA}+\inp{\qq}{\kB}-\inp{\kB}{\kA})$} and  \formula{$(\inp{\pp}{\kA})/(\inp{\pp}{\kA}+\inp{\pp}{\kB}-\inp{\kB}{\kA})$} are more complicated than in the previously discussed cases. 
Also the terms contributing to the running of the coupling constant would require a more sophisticated treatment.
%

\section{Summary\label{SecSummary}}
In this paper we have analyzed  different forms of the exact tree level QCD spin amplitude for the process $q\bar{q}\to{}\JJ{}gg$ where $\JJ$ may represent any, color singlet, current.
We have found quite well structured representations  consisting of sums and  products of compact gauge invariant parts.
Some of those parts have a well defined meaning, as dominant contributions necessary to define evolution kernels. 
In contrast to previous  studies for QED, the present expressions do not seem to be unique.
This is understandable, since amplitudes and the structure of singularities are more complex in QCD.
In particular, the discussion of singularities for processes where incoming quarks would be interchanged with outgoing gluons should be included in our considerations to constrain ambiguities.

We have used two types of organizations of parts.
The first one, discussed in \Section{SecCommAnti}, is rather directly obtained from Feynman rules, with only a limited set of simplifying algebraic manipulations.
This form can be useful, because it manifests the relation between QCD and QED amplitudes. 
In \Section{SecSingProd}, our attention turned towards a form useful for QCD phenomenology.
Our final expression consist of a sum of 10 gauge invariant parts, each just one line long.
Even at the level of the exact spin amplitude, one could attempt to provide a physical interpretation for these parts.
We have provided such an attempt in \Section{cztery-cztery}.
The results look encouraging, and provide a separation into terms possible to interpret either as responsible for consecutive emissions of gluons or as contribution to the decay of a virtual gluon (running of the coupling constant), they however are superficial to a certain degree.
In approximations, a migration of terms takes place between the gauge invariant parts as organized in the exact amplitude.
Such a phenomenon should be expected and it is probably related to  the choice of scales for the running coupling as accommodated by different schemes of parton shower models \cite{Amati:1980ch,Sterman:1986aj}.

In \Section{SecCases}, we discussed approximations for the amplitude, which consist of BFKL or DGLAP pictures, valid in the appropriate phase space regions. 
As a consequence, some parts of our expressions for the exact amplitude could be dropped and even shorter expressions appeared.
They are straightforward to interpret. At the same time explicit expressions  for the difference with the exact amplitude at any point of phase space exist.
Such correction is straightforward to manipulate in any numerical application, for example in parton shower Monte Carlo applications, using positively defined weights. 

The discussion of simplification presented here is by far incomplete with respect to other discussions present in the literature.
For example, we have not performed partial integration of the phase space variables nor have we introduced intermediate effective quark fields into the description.
In particular, our discussion covered the CCFM picture only marginally, even though we do not neglect any aspects related to coherence. 
%

The main analytical results are the compact expressions for the amplitude given in \Section{SecCommAnti} and \Section{MixedR}.
The discussion of approximations is fragmentary only, however, the localization of gauge invariant parts of the amplitude proved to be useful for that purpose.
For a more complete discussion, we need to collect amplitudes for at least all processes of four external color fields accompanying the color-neutral current.
%
%


\providecommand{\href}[2]{#2}\begingroup\raggedright\endgroup

\end{document}